\documentclass[preprint,10.5pt,authoryear]{elsarticle}
\usepackage{geometry}
\usepackage{amssymb}
\usepackage{amsmath}

\usepackage{amsthm}
\theoremstyle{definition}
\newtheorem{definition}{Definition}
\theoremstyle{plain}
\newtheorem{theorem}{Theorem}
\newtheorem{proposition}{Proposition}
\newtheorem{lemma}{Lemma}

\usepackage{subcaption}  
\usepackage{graphicx}    

\usepackage{booktabs}
\usepackage{multirow}
\usepackage{booktabs}  
\usepackage{enumitem}

\usepackage{etoolbox}

\usepackage{caption}
\usepackage{caption}
\makeatletter
\renewenvironment{proof}[1][\proofname]{%
	\par\pushQED{\qed}%
	\normalfont 
	\topsep6\p@\@plus6\p@\relax
	\trivlist
	\item[\hskip\labelsep
	\bfseries #1\@addpunct{.}]%
}{%
	\popQED\endtrivlist\@endpefalse
}
\makeatother

\begin{document}
	
\begin{frontmatter}

\title{Differential Distance Correlation and Its Applications}

\author[a]{Yixiao Liu\corref{cor1}}
\author[b]{Pengjian Shang}

\cortext[cor1]{Corresponding author. E-mail: 23121729@bjtu.edu.cn}
\address[a]{School of Mathematics and Statistics, Beijing Jiaotong University, Beijing 100044, PR China}
\address[b]{School of Mathematics and Statistics, Beijing Jiaotong University, Beijing 100044, PR China}
		
\begin{abstract}
In this paper, we propose a novel Euclidean-distance-based coefficient, named differential distance correlation, to measure the strength of dependence between a random variable \( Y \in \mathbb{R} \) and a random vector \( \boldsymbol{X} \in \mathbb{R}^p \). The coefficient has a concise expression and is invariant to arbitrary orthogonal transformations of the random vector. Moreover, the coefficient is a strongly consistent estimator of a simple and interpretable dependent measure, which is 0 if and only if \( \boldsymbol{X} \) and \( Y \) are independent and equal to 1 if and only if \( Y \) determines \( \boldsymbol{X} \) almost surely. An alternative approach is also proposed to address the limitation that the coefficient is non-robust to outliers. Furthermore, the coefficient exhibits asymptotic normality with a simple variance under the independent hypothesis, facilitating fast and accurate estimation of \( p \)-value for testing independence. Three simulation experiments show that the proposed coefficient is more computationally efficient for independence testing and more effective in detecting oscillatory relationships than several competing methods. We also apply our method to analyze a real data example.
\end{abstract}

		
\end{frontmatter}



\section{Introduction}\label{introduction}
It is fundamental to detect the dependence between random variables in statistical research. One of the most famous dependence measures is the distance correlation \citep{szekely2007}, and \citet{sejdinovic2013} proved that it is equivalent to another dependence measure, the Hilbert-Schmidt independence criterion (HSIC) \citep{gretton2005}. Inspired by the introduction of distance correlation, various dependence measures were developed. For instance, \citet{zhu2017} proposed the projection correlation that does not need moment limitations, and the improved projection correlation with lower computational cost was proposed subsequently \citep{zhang2023}. Moreover, \citet{zhang2025} considered the projection metric in reproducing Hilbert kernel space. Besides, \citet{pan2020} developed the ball correlation to describe the dependence in Banach space, \citet{bergsma2014} proposed the Tau-star coefficient for measuring the dependence between two random variables, and \citet{kong2019} mentioned a composite coefficient of determination. These metrics are equal to zero if and only if the two variables are independent.

However, these methods are subject to some limitations. First, these coefficients sometimes fail to measure the strength of the relationship between random variables. For example, for random variable \( X \) and \( Y = f(X) \) where \( f(\cdot) \) is an intricate function, these aforementioned coefficients between them might be small, but the two variables have a strong association. Second, these metrics do not have simple asymptotic theories under the hypothesis of independence, which makes it difficult to calculate \( p \)-value. Although we can approximate \( p \)-value through random permutation, this approach requires substantial computational cost. Besides, some correlation coefficients require additional choices. Taking HSIC as an example, the performance of its independence test critically depends on the selection of kernel function. Furthermore, the estimation of some metrics requires prohibitive computational cost. For instance, the estimation of projection correlation has \( O(n^3) \) computational cost, where \( n \) represents the sample size. Such high computational costs present significant challenges for tasks like independence testing and feature selection based on these measures. To solve these problems, \citet{chatterjee2021} developed a novel rank-based measure that could measure the functional dependence between two random variables. Let \((X,Y)\) be a pair of random variables, where \( Y \) is not a constant, and \(\{(X_i,Y_j)\}_{i=1}^n\) be the independent samples from \((X,Y)\). Rearrange these samples as \(\{(X_{(i)},Y_{(i)})\}_{i=1}^n\) such that \( X_{(1)} \leq \cdots \leq X_{(n)} \). Define \( r_i \) and \( l_i \) to be the number of \( k \) such that \( Y_{(k)} \leq Y_{(i)} \) and \( Y_{(k)} \geq Y_{(i)} \), respectively. The Chatterjee's coefficient is defined as:

\begin{equation}\label{eq:chatterjee}
	\xi_n(X,Y) := 1 - \frac{n\sum_{i=1}^{n-1}|r_{i+1}-r_i|}{2\sum_{i=1}^n l_i(n-l_i)}.
\end{equation}

The coefficient consistently estimates a simple and interpretable measure of dependence between random variables, which is zero if and only if the two variables are independent and is equal to one if and only if \( Y \) is a measurable function of \( X \) almost surely. Moreover, the coefficient has asymptotic normality under the hypothesis of independence, and the computational cost of it is \( O(n\log n) \). Furthermore, \citet{shi2021} pointed out that the independence testing by Chatterjee's coefficient exhibits lower power compared to some classical measures. Additionally, \citet{cao2020} and \citet{bickel2022} also analyzed its testing power. \citet{lin2022} further proposed an alternative method to enhance its power, essentially by using multiple right-nearest-neighbors instead of the single neighbor adopted in Chatterjee's coefficient. However, this coefficient has two obvious limitations. Firstly, it is restricted to measuring the association between univariate random variables. Secondly, as a rank-based coefficient, it ignores some critical information contained in the original data space, such as the metric differences between observations.

To address these limitations, in this paper, we propose the differential distance correlation, a novel coefficient that measures the strength of the relationship between a random vector \( \boldsymbol{X} \in \mathbb{R}^p \) and a random variable \( Y \in \mathbb{R} \). The coefficient has a simple form. For samples \(\{(\boldsymbol{X}_i,Y_i)\}_{i=1}^n\), it only requires Euclidean distances between \(\{\boldsymbol{X}_i\}_{i=1}^n\) and the rank of \(\{Y_i\}_{i=1}^n\). Notably, when \( X \) is univariate, the computational cost of it is \( O(n \log n) \). Besides, when the first moment of \( \boldsymbol{X} \) exists, the coefficient converges to a simple and interpretable measure of dependence between the two variables almost surely, which is 0 if and only if the variables are independent and 1 if and only if there exists a measurable function vector \( \boldsymbol{f} = (f_1,\ldots,f_p): \mathbb{R} \to \mathbb{R}^p \) such that \( \boldsymbol{f}(Y) = \boldsymbol{X} \) (i.e., \( Y \) determines \( \boldsymbol{X} \) almost surely). In addition, under the conditions that (i) the fourth moment of \( \boldsymbol{X} \) exists and (ii) \( \boldsymbol{X} \) and \( Y \) are independent, the proposed coefficient exhibits asymptotic normality with an easily estimable variance. Moreover, a simple approach that not only avoids the moment limitations on \( \boldsymbol{X} \) but also preserves the advantages of the coefficient is proposed. Three simulated experiments are then conducted to illustrate the properties of our proposed coefficient in several tasks. Furthermore, we apply our coefficient to select relevant predictors from a microarray dataset.

This paper is organized as follows. In Section 2, we present the definition and several statistical properties of the differential distance correlation and the measure it consistently estimates. Furthermore, we establish its asymptotic normality under the independence hypothesis and introduce an alternative method that avoids the moment limitations. In Section 3, we conduct three simulated experiments and a real data example. In Section 4, we summarize the limitations of our method and outline several possible directions for future research. Finally, we summarize our work in Section 5. Some technical proofs are provided in the Appendix.

\section{Differential Distance Correlation}
\label{sec:ddc}

\subsection{Definition and Properties}
\label{subsec:definition}

Let \( \boldsymbol{X} \) be a random vector in \( \mathbb{R}^p \), \( Y \) be a random variable in \( \mathbb{R} \), and denote \( \langle \boldsymbol{a}, \boldsymbol{b} \rangle \) as the Euclidean inner product of \( \boldsymbol{a} \) and \( \boldsymbol{b} \). It is natural that \( \boldsymbol{X} \) and \( Y \) are independent if and only if
\begin{equation}
	\label{eq:independence_condition}
	E \left[ e^{i\langle \boldsymbol{t},\boldsymbol{X} \rangle} | Y \right] = E \left[ e^{i\langle \boldsymbol{t},\boldsymbol{X} \rangle} \right] , \boldsymbol{t} \in \mathbb{R}^p,
\end{equation}
where \( i \) is the imaginary number defined by \( i^2 = -1 \). \eqref{eq:independence_condition} holds if and only if
\begin{equation}
	\label{eq:integral_condition}
	\int_{\mathbb{R}^p} E \left[ \left| E \left[ e^{i\langle \boldsymbol{t},\boldsymbol{X} \rangle} | Y \right] - E \left[ e^{i\langle \boldsymbol{t},\boldsymbol{X} \rangle} \right] \right|^2 \right] w(\boldsymbol{t}) dt = 0,
\end{equation}
where \( w(\boldsymbol{t}) \) is an arbitrary positive weight function for which the integral above exists. Here we choose the weight function \( w(\boldsymbol{t}) = (c_p \| \boldsymbol{t} \|^{p+1})^{-1} \) adopted in the distance correlation \citep{szekely2007}, where \(\| \bullet \|\) is the Euclidean norm, \( c_p = \pi^{(1 + p)/2} / \Gamma \left( (1 + p)/2 \right) \), and \(\Gamma (\cdot)\) is gamma function. If \( E[\| \boldsymbol{X} \|] < + \infty \), by the Fubini's theorem and Lemma 1 in \citet{szekely2007}, the left side of \eqref{eq:integral_condition} can be simplified as follows:

\begin{align}
	 \int_{\mathbb{R}^p} &E \left[ \left| E \left[ e^{i\langle \boldsymbol{t},\boldsymbol{X} \rangle} | Y \right] - E \left[ e^{i\langle \boldsymbol{t},\boldsymbol{X} \rangle} \right] \right|^2 \right] w(\boldsymbol{t}) d\boldsymbol{t} \nonumber \\
	&= \int_{\mathbb{R}^p} \left[\left( 1 - \left| E \left[ e^{i\langle \boldsymbol{t},\boldsymbol{X} \rangle} \right] \right|^2 \right)- \left( 1 - E \left[ \left| E \left[ e^{i\langle \boldsymbol{t},\boldsymbol{X} \rangle} | Y \right] \right|^2 \right] \right)\right] w(\boldsymbol{t}) d\boldsymbol{t} \nonumber \\
	&= \int_{\mathbb{R}^p} \left[\left( 1 - E \left[ \cos \langle \boldsymbol{t}, \boldsymbol{X}_1 - \boldsymbol{X}_2 \rangle \right] \right)- \left( 1 - E_{Y} \left[ E \left[ \cos \langle \boldsymbol{t}, \boldsymbol{X}_1 - \boldsymbol{X}_2 \rangle | Y_1 = Y_2 = Y \right] \right] \right)\right] w(\boldsymbol{t}) d\boldsymbol{t} \nonumber \\
	&= E \left[ \int_{\mathbb{R}^p} \left( 1 - \cos \langle \boldsymbol{t}, \boldsymbol{X}_1 - \boldsymbol{X}_2 \rangle \right) w(\boldsymbol{t}) d\boldsymbol{t} \right] \nonumber - E_{Y} \left[ E \left[ \int_{\mathbb{R}^p} \left( 1 - \cos \langle \boldsymbol{t}, \boldsymbol{X}_1 - \boldsymbol{X}_2 \rangle \right) w(\boldsymbol{t}) d\boldsymbol{t} | Y_1 = Y_2 = Y \right] \right] \nonumber \\
	&= E \left[ \| \boldsymbol{X}_1 - \boldsymbol{X}_2 \| \right] - E_{Y} \left[ E \left[ \| \boldsymbol{X}_1 - \boldsymbol{X}_2 \| | Y_1 = Y_2 = Y \right] \right],
	\label{eq:simplified_integral}
\end{align}
where \( (\boldsymbol{X}_1, Y_1) \) and \( (\boldsymbol{X}_2, Y_2) \) are independent copies of \( (\boldsymbol{X}, Y) \). The result in \eqref{eq:simplified_integral} motivates us to define two critical statistics in Definition~\ref{def:ddc}.

\begin{definition}
	\label{def:ddc}
	For non-degenerate random vector \( \boldsymbol{X} \in \mathbb{R}^p \) satisfying \( E[\| \boldsymbol{X} \|] < + \infty \) and random variable \( Y \in \mathbb{R} \), \( DDC(\boldsymbol{X}, Y) \) is defined as follows:
	\begin{equation}
		\label{eq:ddc_population}
		DDC(\boldsymbol{X}, Y) = 1 - \frac{E_{Y} \left[ E \left[ \| \boldsymbol{X}_1 - \boldsymbol{X}_2 \| | Y_1 = Y_2 = Y \right] \right]}{E[\| \boldsymbol{X}_1 - \boldsymbol{X}_2 \|]},
	\end{equation}
	where \( E[\| \boldsymbol{X}_1 - \boldsymbol{X}_2 \|] \), denoted by \( \Delta \), is the Gini mean difference of \( \boldsymbol{X} \). For the independent samples \( \{(\boldsymbol{X}_i, Y_i)\}_{i=1}^n \) generated from \( (\boldsymbol{X}, Y) \), we rearrange them as \( \{(\boldsymbol{X}_{(i)}, Y_{(i)})\}_{i=1}^n \) satisfying \( Y_{(1)} \leq \cdots \leq Y_{(n)} \). The estimation of \( \Delta \) and the differential distance correlation \( DDC_n(\boldsymbol{X}, Y) \) are defined by
	\begin{equation}
		\label{eq:delta_hat}
		\hat{\Delta}_n = \binom{n}{2}^{-1} \sum_{i>j}^{n} \| \boldsymbol{X}_i - \boldsymbol{X}_j \|,
	\end{equation}
	\begin{equation}
		\label{eq:ddc_sample}
		DDC_n(\boldsymbol{X}, Y) = 1 - \frac{\frac{1}{n-1} \sum_{i=1}^{n-1} \left\| \boldsymbol{X}_{(i)} - \boldsymbol{X}_{(i+1)} \right\|}{\displaystyle\binom{n}{2}^{-1}\sum\limits_{i>j}^n \|\boldsymbol{X}_i - \boldsymbol{X}_j\|} = 1 - \frac{n \sum_{i=1}^{n-1} \left\| \boldsymbol{X}_{(i)} - \boldsymbol{X}_{(i+1)} \right\|}{\sum_{i,j=1}^{n} \left\| \boldsymbol{X}_i - \boldsymbol{X}_j \right\|}.
	\end{equation}
	If all \( \boldsymbol{X}_i \) are equal to a same constant vector, we set \( DDC_n(\boldsymbol{X}, Y) = 0 \).
\end{definition}

By the definitions in \eqref{eq:ddc_population} and \eqref{eq:ddc_sample}, \( DDC_n(\boldsymbol{X}, Y) \) can be used to estimate \( DDC(\boldsymbol{X}, Y) \) since \( Y_{(i)} \) is likely to be close to \( Y_{(i+1)} \). The following theorem summarizes several properties of \( DDC(\boldsymbol{X}, Y) \) and \( DDC_n(\boldsymbol{X}, Y) \), and their proofs are shown in the Appendix.

\begin{theorem}
	\label{thm:ddc_properties}
	For non-degenerate random vector \( \boldsymbol{X} \in \mathbb{R}^p \) such that \( E(\|\boldsymbol{X}\|) < +\infty \), random variable \( Y \in \mathbb{R} \), and their independent observations \(\{(\boldsymbol{X}_i,Y_i)\}_{i=1}^n\), \( DDC(\boldsymbol{X},Y) \) and \( DDC_n(\boldsymbol{X},Y) \) have the following properties:  
	\begin{enumerate}[label=(\roman*)]\normalfont
		\item In general, \( 0 \leq DDC(\boldsymbol{X},Y) \leq 1 \). Particularly, \( DDC(\boldsymbol{X},Y) = 0 \) if and only if \( \boldsymbol{X} \) and \( Y \) are independent, and \( DDC(\boldsymbol{X},Y) = 1 \) if and only if there exists a measurable function vector \( \boldsymbol{f} = (f_1,\ldots,f_p): \mathbb{R} \to \mathbb{R}^p \) such that \( \boldsymbol{f}(Y) = \boldsymbol{X} \) almost surely.
		\item \( DDC(a\boldsymbol{C}\boldsymbol{X} + \boldsymbol{b},h(Y)) = DDC(\boldsymbol{X},Y) \) and \( DDC_n(a\boldsymbol{C}\boldsymbol{X} + \boldsymbol{b},h(Y)) = DDC_n(\boldsymbol{X},Y) \) for any nonzero constant \( a \), orthonormal matrix \( \boldsymbol{C} \in \mathbb{R}^{p \times p} \), vector \( \boldsymbol{b} \in \mathbb{R}^p \), and strictly monotonic function \( h: \mathbb{R} \to \mathbb{R} \).
		\item If \( Y \) is not almost surely a constant, \( DDC_n(\boldsymbol{X},Y) \) converges to \( DDC(\boldsymbol{X},Y) \) almost surely as \( n \to \infty \).
		\item If \( (X,Y) \) are bivariate standard normal with correlation coefficient \( \rho \), \( DDC(X,Y) = DDC(Y,X) = 1 - \sqrt{1 - \rho^2} \).
	\end{enumerate}
\end{theorem}

The computational cost of estimating the differential distance correlation is generally \( O(n^2) \). However, this cost can be significantly reduced to \( O(n \log n) \) when \( X \) is univariate. This reduction is achieved by sorting the sample. Let \( X^{(1)} \leq \cdots \leq X^{(n)} \) denote the order statistics of the univariate sample \(\{X_i\}_{i=1}^n \), then the estimation of \( \Delta \) can be simplified as:
\begin{equation}
	\label{eq:delta_univariate}
	\hat{\Delta}_n = \binom{n}{2}^{-1} \sum_{i>j}^n \left\| X_i - X_j \right\| = \binom{n}{2}^{-1} \sum_{i>j}^n (X^{(i)} - X^{(j)}) = \binom{n}{2}^{-1} \sum_{i=1}^n (2i - n - 1)X^{(i)}.
\end{equation}

\subsection{Another Explanation of \( DDC(\boldsymbol{X},Y) \)}
\label{subsec:alternative_explanation}

Another approach for constructing \( DDC(\boldsymbol{X},Y) \) is introduced in this section. It is well-known that \( \boldsymbol{X} \) and \( Y \) are independent if and only if \( \langle \boldsymbol{\alpha}, \boldsymbol{X} \rangle \) and \( Y \) are independent for any vector \( \boldsymbol{\alpha} \in \mathbb{R}^p \). Therefore, testing whether \( \boldsymbol{X} \) and \( Y \) are independent is equivalent to testing whether

\begin{equation}
	\label{eq:independence_test}
	E_{Y} \left[ \int_{\mathbb{R}} \int_{\mathbb{R}^p} \left( F_{\langle \boldsymbol{\alpha}, \boldsymbol{X} \rangle}(s) - F_{\langle \boldsymbol{\alpha}, \boldsymbol{X} \rangle|Y}(s) \right)^2 \varphi(\boldsymbol{\alpha}) d\boldsymbol{\alpha} ds \right] = 0,
\end{equation}
where \( \varphi(\boldsymbol{\alpha}) \) is the standard normal density, \( F_{\langle \boldsymbol{\alpha}, \boldsymbol{X} \rangle}(s) \) and \( F_{\langle \boldsymbol{\alpha}, \boldsymbol{X} \rangle|Y}(s) \) represent the cumulative distribution function and conditional cumulative distribution function of \( \langle \boldsymbol{\alpha}, \boldsymbol{X} \rangle \) given \( Y \), respectively. The left side of \eqref{eq:independence_test} can be represented in the form of indicator function as follows:

\begin{equation}
	\label{eq:indicator_representation}
	\int_{\mathbb{R}} \int_{\mathbb{R}^{p}} \text{Var}\bigl{(}E(I(\langle \boldsymbol{\alpha}, \boldsymbol{X} \rangle \leq s)|Y)\bigr{)} \varphi(\boldsymbol{\alpha}) \, d\boldsymbol{\alpha} ds.
\end{equation}

The following theorem shows an alternative expression for \( DDC(\boldsymbol{X},Y) \).

\begin{theorem}
	\label{thm:ddc_alternative}
	The \( DDC(\boldsymbol{X},Y) \) can be represented as:
	\begin{equation}
		\label{eq:ddc_alternative}
		DDC(\boldsymbol{X},Y) = \frac{\int_{\mathbb{R}} \int_{\mathbb{R}^{p}} \text{Var}\bigl{(}E(I(\langle \boldsymbol{\alpha}, \boldsymbol{X} \rangle \leq s)|Y)\bigr{)} \varphi(\boldsymbol{\alpha}) \, d\boldsymbol{\alpha} ds}{\int_{\mathbb{R}} \int_{\mathbb{R}^{p}} \text{Var}\bigl{(}I(\langle \boldsymbol{\alpha}, \boldsymbol{X} \rangle \leq s)\bigr{)} \varphi(\boldsymbol{\alpha}) \, d\boldsymbol{\alpha} ds}.
	\end{equation}
\end{theorem}

\begin{proof}
	By the law of total variance shown below:
	\[
	\text{Var}(X) = E[\text{Var}(X|Y)] + \text{Var}[E(X|Y)],
	\]
	the numerator of \eqref{eq:ddc_alternative} can be divided into the following two parts:
	\begin{align*}
		 \int_{\mathbb{R}} \int_{\mathbb{R}^{p}} \text{Var}\bigl{(}E(I(\langle \boldsymbol{\alpha}, \boldsymbol{X} \rangle \leq s)|Y)\bigr{)} \varphi(\boldsymbol{\alpha}) \, d\boldsymbol{\alpha} ds 
		= \int_{\mathbb{R}} \int_{\mathbb{R}^{p}} \text{Var}\bigl{(}I(\langle \boldsymbol{\alpha}, \boldsymbol{X} \rangle \leq s)\bigr{)} \varphi(\boldsymbol{\alpha}) \, d\boldsymbol{\alpha} ds 
		 \quad \\
		 - \int_{\mathbb{R}} \int_{\mathbb{R}^{p}} E[\text{Var}(I(\langle \boldsymbol{\alpha}, \boldsymbol{X} \rangle \leq s)|Y)] \varphi(\boldsymbol{\alpha}) \, d\boldsymbol{\alpha} ds 
		= S_{1}-S_{2} ,
	\end{align*}
	where \( S_{1} \) and \( S_{2} \) are defined in the obvious manner. The two parts can be simplified as follows:
	\begin{align*}
		S_{1} &= \frac{1}{2} \int_{\mathbb{R}} \int_{\mathbb{R}^{p}} E\left[|I(\langle \boldsymbol{\alpha}, \boldsymbol{X}_1 \rangle \leq s)-I(\langle \boldsymbol{\alpha}, \boldsymbol{X}_2 \rangle \leq s)|\right] \varphi(\boldsymbol{\alpha}) \, d\boldsymbol{\alpha} ds \\
		&= \frac{1}{2} \int_{\mathbb{R}^{p}} E\left[|\langle \boldsymbol{\alpha}, \boldsymbol{X}_1-\boldsymbol{X}_2 \rangle|\right] \varphi(\boldsymbol{\alpha}) \, d\boldsymbol{\alpha} = \frac{1}{\sqrt{2\pi}} E[\|\boldsymbol{X}_1-\boldsymbol{X}_2\|], \\
		S_{2} &= \frac{1}{2} E_{Y} \left[ \int_{\mathbb{R}} \int_{\mathbb{R}^{p}} E\left[|I(\langle \boldsymbol{\alpha}, \boldsymbol{X}_1 \rangle \leq s)-I(\langle \boldsymbol{\alpha}, \boldsymbol{X}_2 \rangle \leq s)||Y_1 = Y_2 = Y\right] \varphi(\boldsymbol{\alpha}) \, d\boldsymbol{\alpha} ds \right] \\
		&= \frac{1}{\sqrt{2\pi}} E_{Y} [E[\|\boldsymbol{X}_1-\boldsymbol{X}_2\||Y_1=Y_2=Y]].
	\end{align*}
	Therefore, \( DDC(\boldsymbol{X},Y) \) can be represented as
	\[
	DDC(\boldsymbol{X},Y) = \frac{S_{1}-S_{2}}{S_{1}}.
	\]
\end{proof}

\subsection{Asymptotic Normality}
\label{subsec:asymptotic_normality}

In this section, we present the asymptotic properties of the differential distance correlation under the independent hypothesis.

\begin{theorem}
	\label{thm:asymptotic_normality}
	For non-degenerate random vector \( \boldsymbol{X} \in \mathbb{R}^{p} \) satisfying \( E[\|\boldsymbol{X}\|^{4}] < +\infty \) and \( Y \in \mathbb{R} \), where \( \boldsymbol{X} \) and \( Y \) are independent, the \( DDC_{n}(\boldsymbol{X},Y) \) has the following asymptotic property:
	\begin{equation}
		\label{eq:asymptotic_normality}
		\sqrt{n} DDC_{n}(\boldsymbol{X},Y) \xrightarrow{D} N\left(0, \frac{(dVar(\boldsymbol{X}))^{2}}{\Delta^{2}}\right),
	\end{equation}
	where \( dVar(\boldsymbol{X}) \) is the distance variance of \( \boldsymbol{X} \) \citep{szekely2007}, and it can be expressed as
	\begin{equation}
		\label{eq:distance_variance}
		(dVar(\boldsymbol{X}))^{2} = E[\|\boldsymbol{X}_1-\boldsymbol{X}_2\|^{2}] - 2E[\|\boldsymbol{X}_1-\boldsymbol{X}_2\|\|\boldsymbol{X}_1-\boldsymbol{X}_3\|] + E[\|\boldsymbol{X}_1-\boldsymbol{X}_2\|]^{2}
	\end{equation}
	since the second moment of \( \boldsymbol{X} \) exists.
\end{theorem}

We denote the asymptotic variance in \eqref{eq:asymptotic_normality} by \( \sigma^{2} \). In particular, \( \sigma^{2} = \pi/3-\sqrt{3}+1 \) when \( X \) follows univariate normal distribution, and it equals \( 2/5 \) when \( X \) follows the uniform distribution. The proofs of Theorem~\ref{thm:asymptotic_normality} and the two cases are provided in the Appendix.

Then, we discuss the estimation of the asymptotic variance. First, \( \hat{\Delta}_{n} \) converges to \( \Delta \) almost surely as \( n\to\infty \), and the estimation of \( (dVar(\boldsymbol{X}))^{2} \) shown in equation (2.9) of \citet{szekely2007}, denoted by \( \mathcal{V}^{2}_{n} \), converges to \( (dVar(\boldsymbol{X}))^{2} \) almost surely as \( n\to\infty \). By the continuous mapping theorem, the estimated variance \( \hat{\sigma}_{n}^{2} = \mathcal{V}^{2}_{n} \big/ \hat{\Delta}_{n}^{2} \) converges to \( \sigma^{2} \) almost surely. Based on the definition of \( \hat{\Delta}_{n} \) and \( \mathcal{V}^{2}_{n} \), the estimated variance has \( O(n^{2}) \) computational cost. However, when \( X \) is univariate, combining the fast algorithm for estimating \( \mathcal{V}^{2}_{n} \) in \citet{chaudhuri2019} and \( \hat{\Delta}_{n} \) in \eqref{eq:delta_univariate}, the computational cost can be reduced to \( O(n\log n) \).

\subsection{The Extension of \( DDC(\boldsymbol{X},Y) \)}
\label{subsec:extension}

In this section, we introduce an alternative method designed to overcome the sensitivity to outliers that arises from the limitation of moments. The following proposition offers the theoretical guarantee for the alternative approach, and its proof is given in the Appendix.

\begin{proposition}
	\label{prop:extension}
	For arbitrary \( p \)-dimensional strictly monotone function vector \( \boldsymbol{h}=(h_{1},\ldots,h_{p}):\mathbb{R}^{p}\to\mathbb{R}^{p} \) such that \( E\big(\|\boldsymbol{h}(\boldsymbol{X})\|\big) < +\infty \),
	\begin{enumerate}[label=(\roman*)]\normalfont
		\item \( DDC(\boldsymbol{h}(\boldsymbol{X}),Y)=0 \) if and only if \( \boldsymbol{X} \) and \( Y \) are independent.
		\item \( DDC(\boldsymbol{h}(\boldsymbol{X}),Y)=1 \) if and only if there exists a measurable function \( \boldsymbol{f}=(f_{1},\ldots,f_{p}):\mathbb{R}\to\mathbb{R}^{p} \) such that \( \boldsymbol{f}(Y)=\boldsymbol{X} \) almost surely.
		\item If $ Y $ is not almost surely a constant, \(  DDC_{n}(\boldsymbol{h}(\boldsymbol{X}),Y) \) converges to \( DDC(\boldsymbol{h}(\boldsymbol{X}),Y) \) almost surely as $ n \to \infty $.
		\item When (a) \( E[\|\boldsymbol{h}(\boldsymbol{X})\|^{4}] < +\infty \) and (b) \( \boldsymbol{X} \) and \( Y \) are independent, \( DDC_{n}(\boldsymbol{h}(\boldsymbol{X}),Y) \) exhibits asymptotic normality.
	\end{enumerate}
\end{proposition}

Based on Proposition~\ref{prop:extension}, appropriate strictly monotonic functions, such as the sigmoid and arc-tangent function, can be used to address the limitation that \( DDC_{n}(\boldsymbol{X},Y) \) is non-robust to outliers, while preserving its ability to measure independence and functional dependence. Notably, when the arc-tangent function \( atan(x) \) is employed and the dimension of \( X \) is 1, \( DDC(atan(X),Y) \) is equal to the trace correlation \citep{zhao2026} since $ |atan(X_{1})-atan(X_{2})|=\text{arccos}\left\{\frac{1 + \langle X_1,X_2 \rangle}{\sqrt{(1 + \langle X_1,X_1 \rangle)(1 + \langle X_2,X_2 \rangle)}}\right\} \ $.

\section{Numerical Experiments}
\label{sec:experiments}

In this section, we conduct three simulated experiments to evaluate the performance of the differential distance correlation for the independence test and its run-time, functional dependence identification, and feature selection. Then, a real-world dataset is employed to demonstrate the efficacy of our method in identifying response-relevant features.

Before conducting the experiments, we first introduce the methods used for comparison. In addition to the Chatterjee's coefficient shown in \eqref{eq:chatterjee}, we also consider other methods: distance correlation (denoted by DC) \citep{szekely2007}, Tau-star coefficient \citep{bergsma2014}, the HHG test \citep{heller2013}, Ball correlation \citep{pan2020}, Hilbert-Schmidt Independence Criterion (denoted by HSIC) \citep{gretton2005} with the Gaussian kernel of the median-distance based bandwidth, and the improved projection correlation (denoted by PCor) \citep{zhang2023} that can be expressed as:
\begin{equation}
	\label{eq:pcor_pcov}
\begin{aligned}
		PCov(\boldsymbol{X},\boldsymbol{Y}) &= E\left[A(\boldsymbol{X_1},\boldsymbol{X_2})A(\boldsymbol{Y_1},\boldsymbol{Y_2}) - 2A(\boldsymbol{X_1},\boldsymbol{X_2})A(\boldsymbol{Y_1},\boldsymbol{Y_3}) + A(\boldsymbol{X_1},\boldsymbol{X_2})A(\boldsymbol{Y_3},\boldsymbol{Y_4})\right], \\
		PCor(\boldsymbol{X},\boldsymbol{Y}) &= \frac{PCov(\boldsymbol{X},\boldsymbol{Y})}{\sqrt{PCov(\boldsymbol{X},\boldsymbol{X})PCov(\boldsymbol{Y},\boldsymbol{Y})}},
\end{aligned}
\end{equation}
where \( A(\boldsymbol{Z_1},\boldsymbol{Z_2}) = \text{arccos}\left\{\frac{\sigma^2 + \langle \boldsymbol{Z_1},\boldsymbol{Z_2} \rangle}{\sqrt{(\sigma^2 + \langle \boldsymbol{Z_1},\boldsymbol{Z_1} \rangle)(\sigma^2 + \langle \boldsymbol{Z_2},\boldsymbol{Z_2} \rangle)}}\right\} \), and we take \( \sigma^2 = 1 \) in these experiments. Besides, we denote our proposed coefficient as its abbreviation: DDC.

\subsection*{Example 1. Run-time comparison of the independence test}

First, we compare the run times of independence tests based on the eight competing test statistics. The R packages \texttt{XICOR}, \texttt{Energy}, \texttt{dHSIC}, \texttt{TauStar}, \texttt{HHG}, and \texttt{Ball} are utilized to conduct independence tests for Chatterjee's coefficient, distance correlation, HSIC, Tau-star, HHG test, and Ball correlation, respectively. As no existing R package is available for the improved projection correlation test, we write the code to report the run time of its independence test.

For DDC and Chatterjee's correlation, independence testing is conducted by exploiting their asymptotic normality, while the other methods adopt permutation tests with the number of permutations fixed at 200. The run times are reported in Table~\ref{tab:runtime}. It can be observed that DDC requires considerably less time compared to the methods that rely on permutation tests. This confirms the computational efficiency of our approach in independence testing.

\begin{table}[htbp]
	\centering
	\caption{Run times (in seconds) for eight independence testing methods across different sample sizes. For DDC and Chatterjee's coefficient, the asymptotic tests are used. The other six methods adopt permutation tests with 200 permutations.}
	\label{tab:runtime}
	\hrule 
	\vspace{2pt}
	\begin{tabular}{c c c c c c c c c}  
		n & DDC & Chatterjee & DC & PCor & HSIC & TauStar & HHG & Ball \\
		\cmidrule(lr){1-1} \cmidrule(lr){2-2} \cmidrule(lr){3-3} \cmidrule(lr){4-4} \cmidrule(lr){5-5} \cmidrule(lr){6-6} \cmidrule(lr){7-7} \cmidrule(lr){8-8} \cmidrule(lr){9-9} 
		200 & 0.005 & 0.001 & 0.005 & 0.005 & 0.142 & 0.062 & 0.326 & 0.048 \\
		500 & 0.011 & 0.001 & 0.049 & 0.068 & 0.623 & 0.373 & 1.211 & 0.279 \\
		1000 & 0.023 & 0.001 & 0.176 & 0.212 & 2.823 & 2.474 & 5.015 & 1.788 \\
		2000 & 0.055 & 0.001 & 1.501 & 1.167 & 11.460 & 22.002 & 22.305 & 10.254 \\
		4000 & 0.106 & 0.002 & 5.142 & 4.742 & 59.490 & 91.079 & 133.974 & 72.911 \\
	\end{tabular}
	\vspace{2pt}
	\hrule 
\end{table}

\subsection*{Example 2. Independence Test and Functional Dependence Identification}\label{relationship}

We consider the following six cases in this experiment, some of which were used in \citet{chatterjee2021}. Generating \(X\) from the uniform distribution on \([-1,1]\), these models are represented as:

\begin{enumerate}[label=\alph*.]
	\item Linear function: \(Y=X+3\lambda\varepsilon\), where \(\lambda\) is the noise level ranging from 0 to 1 and \(\varepsilon\sim N(0,1)\).
	\item Quadratic: \(Y=X^{2}+2\lambda\varepsilon\).
	\item Sinusoid: \(Y=\cos(8\pi X)+3\lambda\varepsilon\).
	\item Damped oscillator: \(Y=e^{-2X}\sin(10X)+4\lambda\varepsilon\).
	\item W-shaped: \(Y=|X+0.5|I(X<0)+|X-0.5|I(X\geq 0)+0.75\lambda\varepsilon\), where \(I\) is the indicator function.
	\item Step function: \(Y=-3I\{-1\leq X<-0.5\}+2I\{-0.5\leq X<0\}-4I\{0\leq X<0.5\}-3I\{0.5\leq X \leq 1\}+10\lambda\varepsilon\).
\end{enumerate}

\begin{figure}[htbp]
	\centering
	
	\begin{minipage}{0.32\textwidth}
		\centering
		\includegraphics[width=\linewidth]{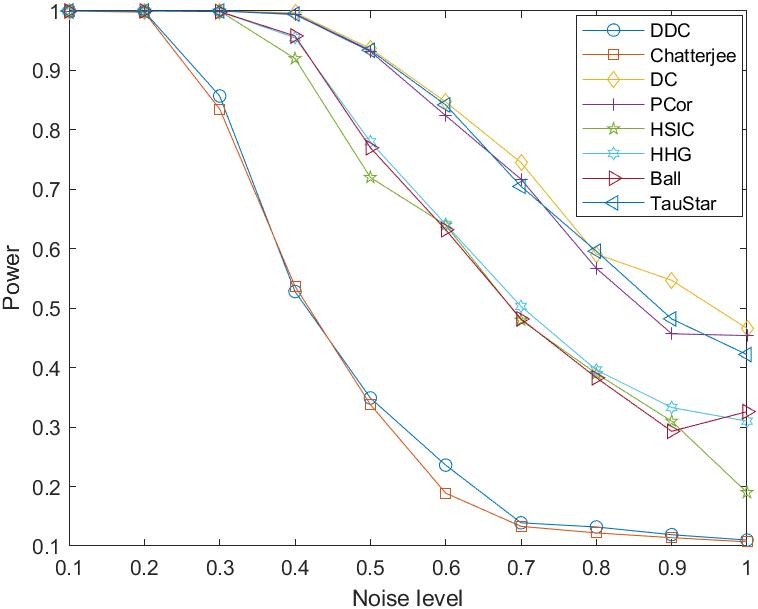}
		\caption*{(a) Linear}
		\label{subfig:linear}
	\end{minipage}
	\hfill
	\begin{minipage}{0.32\textwidth}
		\centering
		\includegraphics[width=\linewidth]{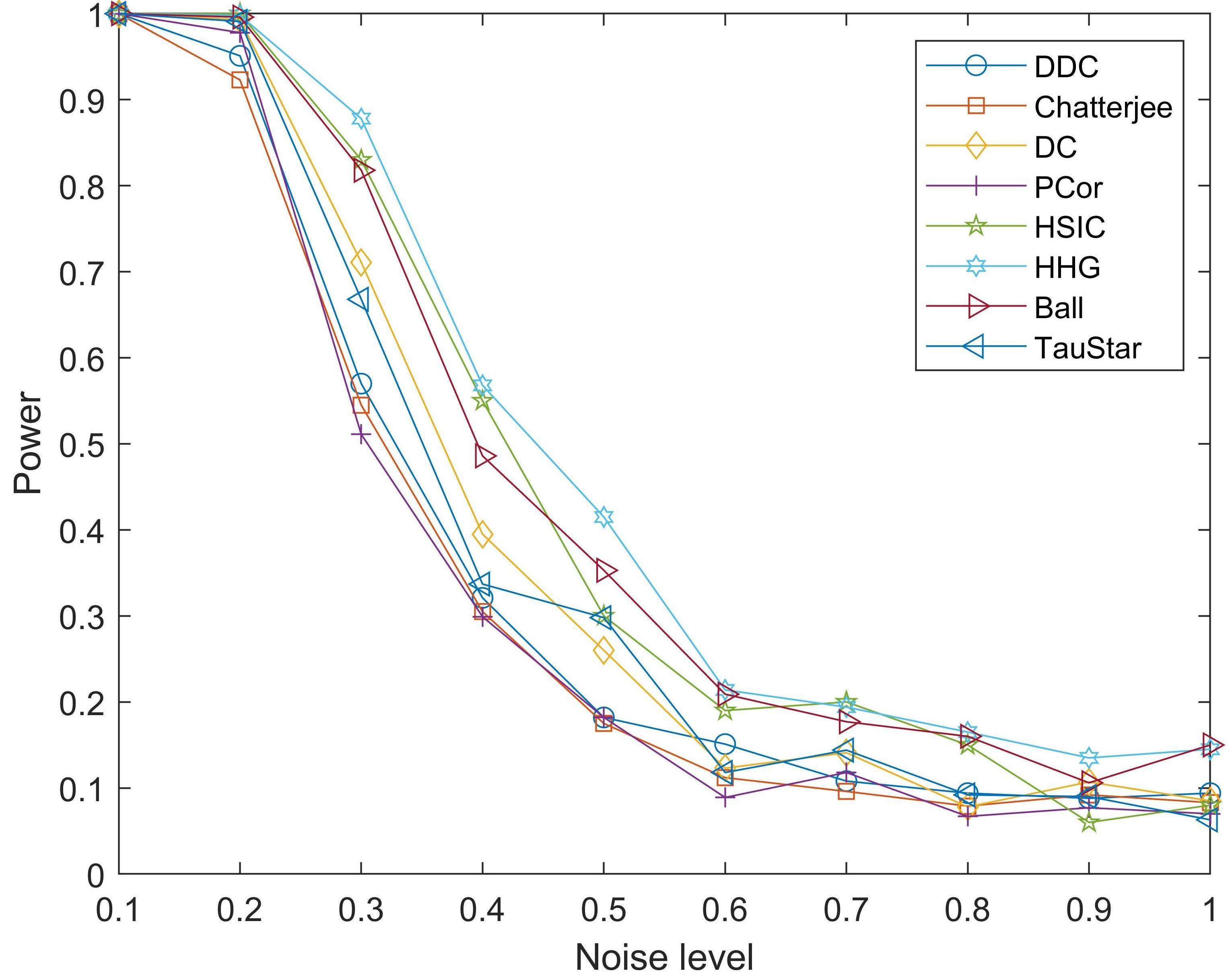}
		\caption*{(b) Quadratic}
		\label{subfig:quadratic}
	\end{minipage}
	\hfill
	\begin{minipage}{0.32\textwidth}
		\centering
		\includegraphics[width=\linewidth]{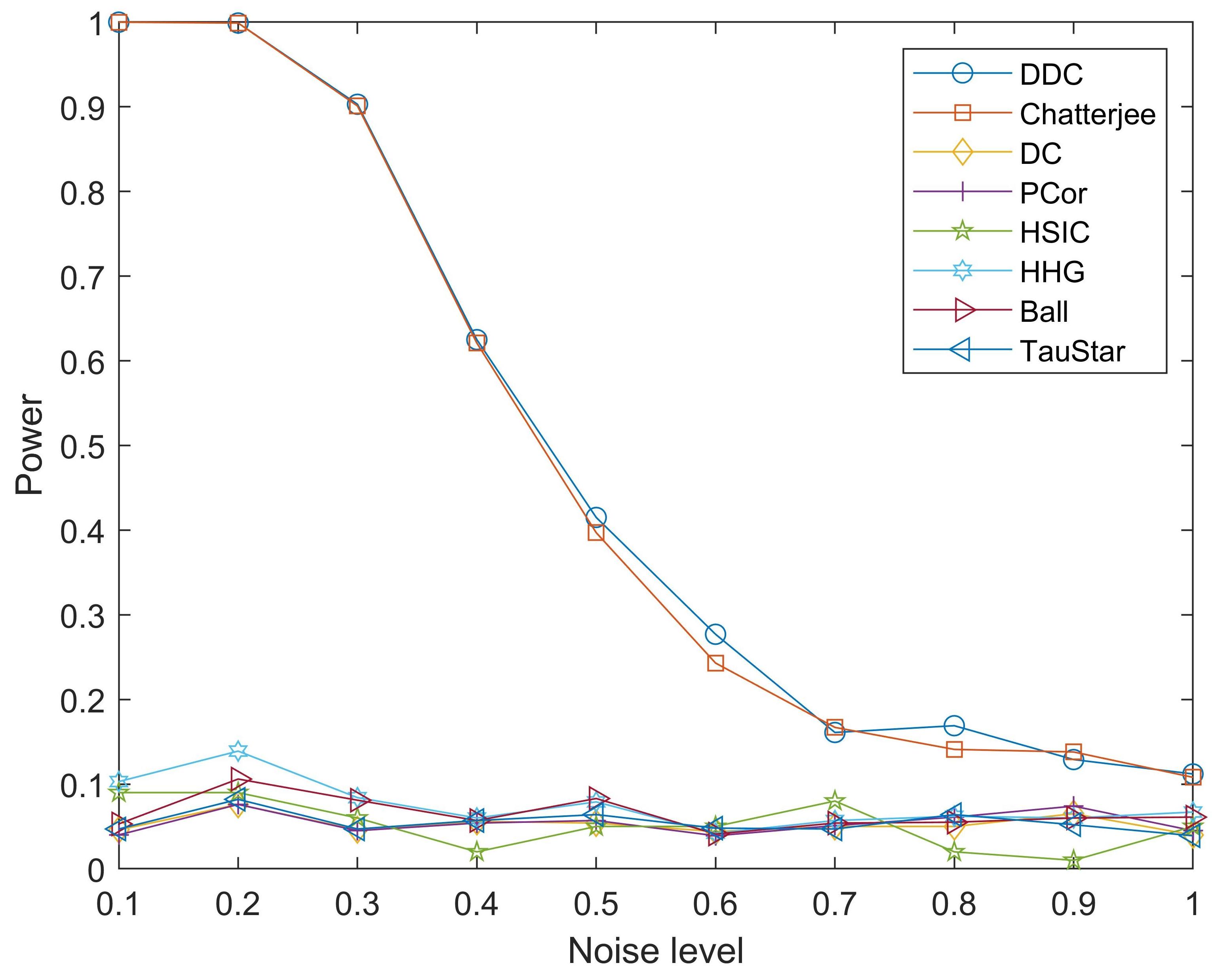}
		\caption*{(c) Sinusoid}
		\label{subfig:sinusoid}
	\end{minipage}
	
	\vspace{0.3cm} 
	
	\begin{minipage}{0.32\textwidth}
		\centering
		\includegraphics[width=\linewidth]{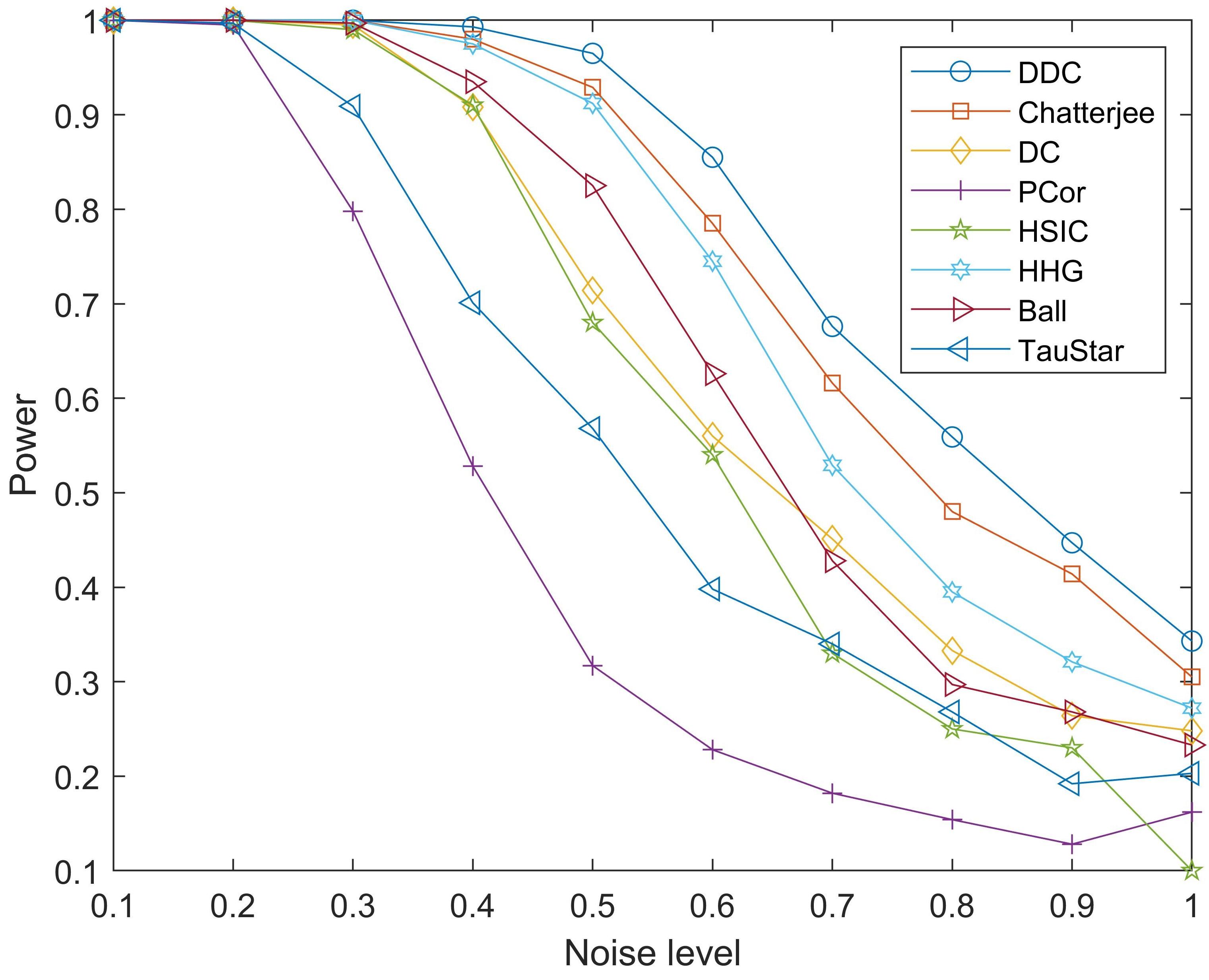}
		\caption*{(d) Damped oscillator}
		\label{subfig:damped}
	\end{minipage}
	\hfill
	\begin{minipage}{0.32\textwidth}
		\centering
		\includegraphics[width=\linewidth]{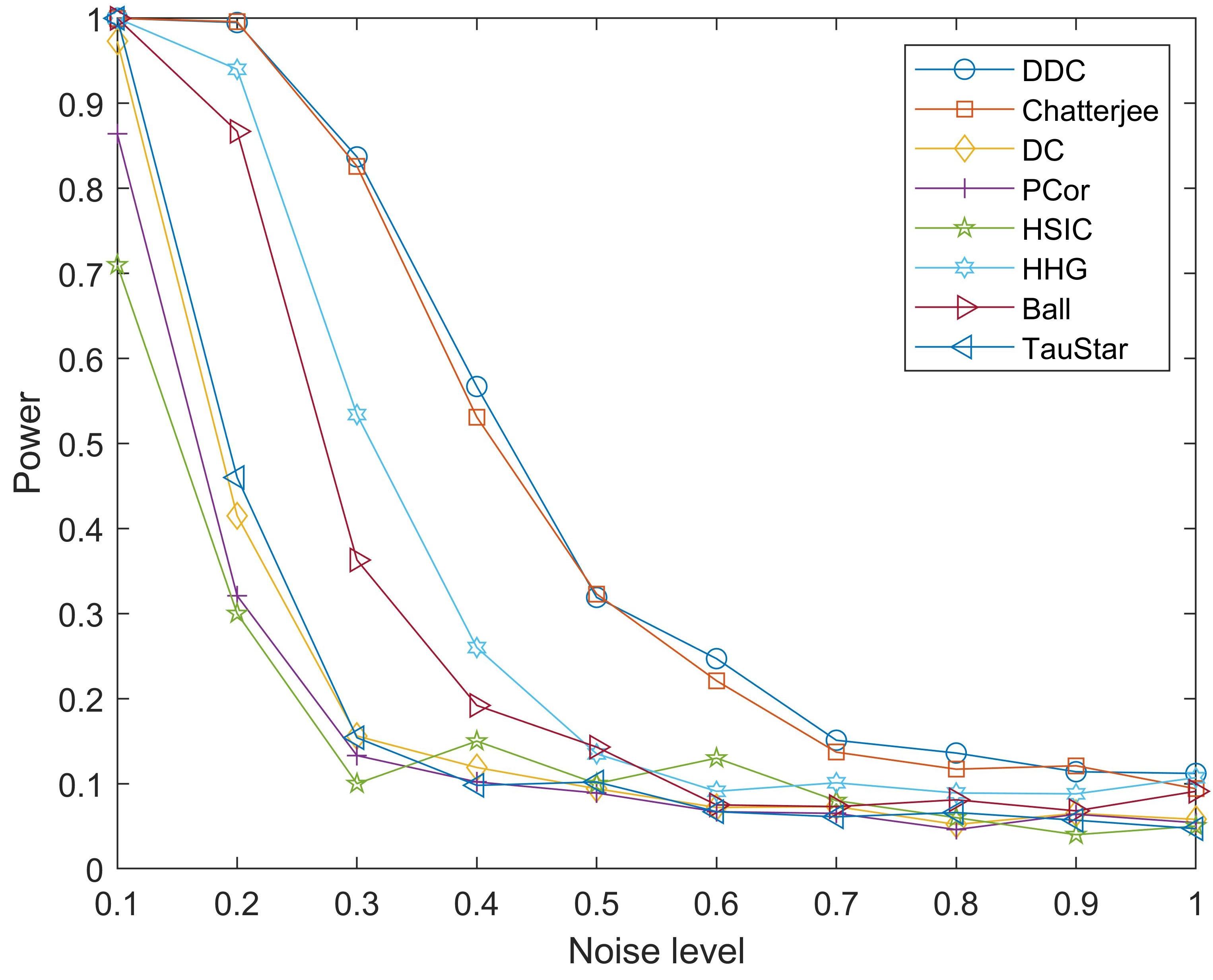}
		\caption*{(e) W-shaped}
		\label{subfig:w_shaped}
	\end{minipage}
	\hfill
	\begin{minipage}{0.32\textwidth}
		\centering
		\includegraphics[width=\linewidth]{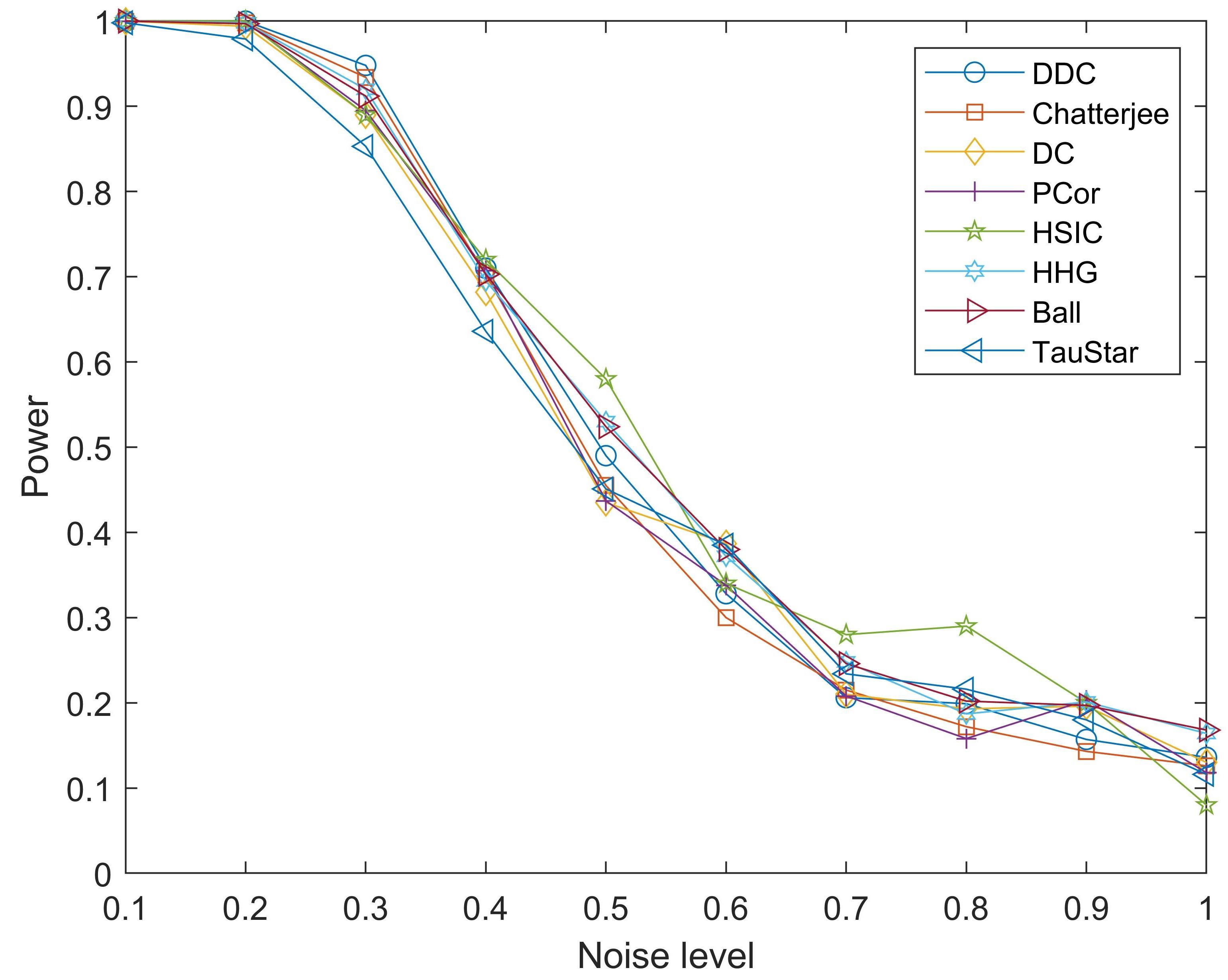}
		\caption*{(f) Step function}
		\label{subfig:step}
	\end{minipage}
	
	\caption{Power comparison of the eight independence testing methods across six functional relationships (linear, quadratic, sinusoid, damped oscillator, W-shaped, step) shown in Example~2 with increasing noise levels. Based on 500 simulations with sample size \(n=100\).}
	\label{fig:power_comparison}
\end{figure}

In each case, the generated sample size is 100, and 500 simulations are used to estimate the power of independence testing. The results are presented in Fig.~\ref{fig:power_comparison}. It can be observed that DDC and Chatterjee's coefficient show similarly independent testing power, except in the damped oscillator case where DDC performs better. Moreover, DDC is more powerful than the other methods when testing oscillatory functions, such as the W-shaped and sinusoid functions. However, DDC demonstrates inferior performance for smoother functions like the linear function.

Furthermore, we explore the strength of functional dependence through the coefficient's value. To ensure comparability, following the improved projection correlation in \eqref{eq:pcor_pcov}, we normalize HSIC using the Cauchy--Schwarz inequality and also square the distance correlation coefficient. Table~\ref{tab:mean_values} shows the mean values of the coefficients from 500 simulations for linear, quadratic, and step-dependent function types and five noise levels. It can be seen that when the noise level is low (i.e., the functional dependence is relatively strong) and the dependence function is fluctuating, both DDC and Chatterjee's coefficient are much higher than the other methods. This indicates that our method, like Chatterjee's coefficient, is capable of detecting dependence, particularly excelling in identifying oscillatory dependencies.

\begin{table}[htbp]
	\centering
	\caption{The mean values of correlation coefficients under different noise levels $\lambda$ and three functional relationships (linear, quadratic, step) shown in Example~2. Experiments are replicated 500 times.}
	\label{tab:mean_values}
	\hrule
	\vspace{2pt}
	\begin{tabular}{c c c c c c c}
		Function & Method & $\lambda=0.1$ & $\lambda=0.3$ & $\lambda=0.5$ & $\lambda=0.7$ & $\lambda=0.9$ \\
		\cmidrule(lr){1-1} \cmidrule(lr){2-2} \cmidrule(lr){3-3} \cmidrule(lr){4-4} \cmidrule(lr){5-5} \cmidrule(lr){6-6} \cmidrule(lr){7-7} 
		\multirow{7}{*}{Linear}
		& DDC & 0.5413 & 0.1588 & 0.0681 & 0.0313 & 0.0244 \\
		& Chatterjee & 0.5576 & 0.1720 & 0.0737 & 0.0335 & 0.0269 \\
		& DC & 0.7705 & 0.2798 & 0.1377 & 0.0850 & 0.0653 \\
		& PCor & 0.7763 & 0.2758 & 0.1293 & 0.0754 & 0.0556 \\
		& HSIC & 0.6457 & 0.1737 & 0.0768 & 0.0508 & 0.0364 \\
		& Ball & 0.2471 & 0.0476 & 0.0238 & 0.0164 & 0.0139 \\
		& TauStar & 0.5561 & 0.1628 & 0.0688 & 0.0355 & 0.0232 \\
		\addlinespace[8pt]
		\multirow{7}{*}{Quadratic}
		& DDC & 0.4376 & 0.1021 & 0.0410 & 0.0181 & 0.0148 \\
		& Chatterjee & 0.4394 & 0.1098 & 0.0436 & 0.0189 & 0.0164 \\
		& DC & 0.1812 & 0.0671 & 0.0449 & 0.0364 & 0.0340 \\
		& PCor & 0.1563 & 0.0571 & 0.0376 & 0.0297 & 0.0269 \\
		& HSIC & 0.2599 & 0.0688 & 0.0385 & 0.0301 & 0.0216 \\
		& Ball & 0.0789 & 0.0209 & 0.0139 & 0.0117 & 0.0108 \\
		& TauStar & 0.1064 & 0.0246 & 0.0103 & 0.0053 & 0.0037 \\
		\addlinespace[8pt]
		\multirow{7}{*}{Step}
		& DDC & 0.5553 & 0.1997 & 0.0903 & 0.0436 & 0.0321 \\
		& Chatterjee & 0.4013 & 0.2078 & 0.0978 & 0.0475 & 0.0351 \\
		& DC & 0.1957 & 0.0976 & 0.0599 & 0.0447 & 0.0389 \\
		& PCor & 0.2074 & 0.1004 & 0.0514 & 0.0348 & 0.0281 \\
		& HSIC & 0.2779 & 0.0994 & 0.0546 & 0.0336 & 0.0284 \\
		& Ball & 0.0558 & 0.0260 & 0.0165 & 0.0131 & 0.0117 \\
		& TauStar & 0.0706 & 0.0400 & 0.0195 & 0.0106 & 0.0067 \\
	\end{tabular}
	\vspace{2pt}
	\hrule
\end{table}

\subsection*{Example 3. Multi-Response Model}\label{multi-response model}

Let's consider a regression model with multiple response variables \( \boldsymbol{Y} = [Y_1, Y_2, Y_3]^T \) shown below:  
\begin{equation}
	\label{eq:multi_response}
	\boldsymbol{Y} = 
	\begin{bmatrix}
		0.2X_1 + 0.2X_2^2 + \sin(4\pi X_3) \\
		0.4X_1 + 0.3X_2^2 + \cos(8\pi X_3) \\
		0.6X_1 - 0.5X_2^2 - \cos(4\pi X_3^2)
	\end{bmatrix} + \boldsymbol{Z}f(X_4) + 0.5\boldsymbol{\varepsilon}.
\end{equation}
The predictor variable \( \boldsymbol{X} = (X_1, ..., X_p)^T \) follows \( N(0, \Sigma) \) with covariance matrix \( \Sigma = (\sigma_{i,j})_{p \times p} \) such that \( \sigma_{i,i} = 1, \, i \in \{1, ..., p\} \) and \( \sigma_{i,j} = \rho, \, i \neq j \). The error term \( \boldsymbol{\varepsilon} \) is independently generated from \( N(0, E_{3 \times 3}) \), where \( E \) represents the identity matrix. $ \boldsymbol{Z} $ is a random vector that takes \([1, 0, 0]^T, \, [0, 1, 0]^T, \, [0, 0, 1]^T\) with equal probability, and \( f(x) = 0.5 \cos(2\pi x) + \cos^2(2\pi x) - 1.5 \sin^3(2\pi x) \). We set \( p = 500 \), generate \( n = 200 \) samples in each simulation, select \( [n/\log n] \) variables as the predictive model, where \( [x] \) represents the integer part of \( x \), and repeat each simulation 100 times. Besides, we consider \( \rho = 0.3, 0.5, \) and 0.7. Table~\ref{tab:proportion} summarizes the proportion of active predictor \( X_i \) in the predictive model (denoted by \( P_i \)), along with the median (MMS) and standard deviation (SD) of the minimal model size containing all active predictors. The results demonstrate that compared to the other measures, our proposed coefficient shows a stronger preference for selecting predictors with oscillatory functional dependence on the response, while exhibiting weaker capability in detecting linearly and quadratic dependent predictors. Furthermore, for all three cases, our method achieves the smallest MMS and SD, indicating more efficiency in selecting all active features. Table~\ref{tab:example3_power} presents the power of different coefficients for testing independence. Our proposed metric remains more powerful in detecting predictors with oscillatory dependent relationships, but is less effective for predictors with smoother functions.

\begin{table}[htbp]
	\centering
	\caption{Selection results for the multi-response model in Example~3 with \(p=500\) predictors, \(n=200\) samples, and 100 simulations. For each value of \(\rho\), columns \(P_1\)--\(P_4\) show the proportion of simulations where true active variables \(X_1\)--\(X_4\) are selected. MMS and SD are the median and standard deviation of the minimal model size containing all four active variables.}
	\label{tab:proportion}
	\hrule
	\vspace{2pt}
	\begin{tabular}{c c c c c c c c}
		$\rho$ & Method & $P_1$ & $P_2$ & $P_3$ & $P_4$ & MMS & SD \\
		\cmidrule(lr){1-1} \cmidrule(lr){2-2} \cmidrule(lr){3-3} \cmidrule(lr){4-4} \cmidrule(lr){5-5} \cmidrule(lr){6-6} \cmidrule(lr){7-7} \cmidrule(lr){8-8}
		\multirow{5}{*}{0.3}
		& DDC & 0.86 & 0.93 & \textbf{1} & \textbf{1} & \textbf{9.5} & \textbf{33.65} \\
		& DC & \textbf{1} & \textbf{1} & 0.08 & 0.22 & 265 & 118.37 \\
		& PCor & \textbf{1} & 0.57 & 0.10 & 0.21 & 262.5 & 112.36 \\
		& HSIC & \textbf{1} & \textbf{1} & 0.14 & 0.05 & 305 & 115.27 \\
		& Ball & \textbf{1} & \textbf{1} & 0.21 & 0.45 & 134 & 99.89 \\
		\addlinespace[8pt]
		\multirow{5}{*}{0.5}
		& DDC & 0.81 & 0.88 & \textbf{1} & \textbf{1} & \textbf{19} & \textbf{67.49} \\
		& DC & \textbf{1} & \textbf{0.96} & 0.09 & 0.13 & 318 & 134.85 \\
		& PCor & \textbf{1} & 0.38 & 0.12 & 0.10 & 313 & 114.21 \\
		& HSIC & \textbf{1} & \textbf{1} & 0.09 & 0.07 & 359 & 123.54 \\
		& Ball & \textbf{1} & \textbf{1} & 0.16 & 0.25 & 220 & 131.26 \\
		\addlinespace[8pt]
		\multirow{5}{*}{0.7}
		& DDC & 0.53 & 0.79 & \textbf{1} & \textbf{1} & \textbf{48.5} & \textbf{86.27} \\
		& DC & \textbf{1} & 0.85 & 0.03 & 0.19 & 321.5 & 120.15 \\
		& PCor & \textbf{1} & 0.28 & 0.07 & 0.17 & 352 & 110.44 \\
		& HSIC & \textbf{1} & 0.95 & 0.06 & 0.10 & 383.5 & 114.29 \\
		& Ball & \textbf{1} & \textbf{0.97} & 0.10 & 0.27 & 256 & 123.28 \\
	\end{tabular}
	\vspace{2pt}
	\hrule
\end{table}

\begin{table}[htbp]
	\centering
	\caption{The independence testing power of different methods for the four active variables \(X_1\), \(X_2\), \(X_3\), \(X_4\) under different $\rho$ in the multi-response model in Example~3. Results are based on 100 simulations.} 
	\label{tab:example3_power}
	\hrule
	\vspace{2pt}
	\begin{tabular}{c c c c c c c c c c c c c}
		 & \multicolumn{4}{c}{\(\rho = 0.3\)} & \multicolumn{4}{c}{\(\rho = 0.5\)} & \multicolumn{4}{c}{\(\rho = 0.7\)} \\
		\cmidrule(lr){2-5} \cmidrule(lr){6-9} \cmidrule(lr){10-13}
		Method & \(X_1\) & \(X_2\) & \(X_3\) & \(X_4\) & \(X_1\) & \(X_2\) & \(X_3\) & \(X_4\) & \(X_1\) & \(X_2\) & \(X_3\) & \(X_4\) \\
		\cmidrule(lr){1-1} \cmidrule(lr){2-2} \cmidrule(lr){3-3} \cmidrule(lr){4-4} \cmidrule(lr){5-5} \cmidrule(lr){6-6} \cmidrule(lr){7-7} \cmidrule(lr){8-8} \cmidrule(lr){9-9} \cmidrule(lr){10-10} \cmidrule(lr){11-11} \cmidrule(lr){12-12} \cmidrule(lr){13-13}
		DDC & 0.88 & 0.97 & \textbf{1} & \textbf{1} & 0.92 & 0.93 & \textbf{1} & \textbf{1} & 0.98 & 0.97 & \textbf{1} & \textbf{1} \\
		DC & \textbf{1} & \textbf{1} & 0.57 & 0.72 & \textbf{1} & \textbf{1} & 0.96 & 0.98 & \textbf{1} & \textbf{1} & \textbf{1} & \textbf{1} \\
		PCor & \textbf{1} & 0.97 & 0.55 & 0.70 & \textbf{1} & \textbf{1} & 0.94 & 0.97 & \textbf{1} & \textbf{1} & \textbf{1} & \textbf{1} \\
		HSIC & \textbf{1} & \textbf{1} & 0.39 & 0.33 & \textbf{1} & \textbf{1} & 0.78 & 0.81 & \textbf{1} & \textbf{1} & \textbf{1} & \textbf{1} \\
		Ball & \textbf{1} & \textbf{1} & 0.59 & 0.81 & \textbf{1} & \textbf{1} & 0.81 & 0.97 & \textbf{1} & \textbf{1} & \textbf{1} & \textbf{1} \\
		HHG & \textbf{1} & \textbf{1} & 0.62 & 0.89 & \textbf{1} & \textbf{1} & 0.89 & 0.98 & \textbf{1} & \textbf{1} & \textbf{1} & \textbf{1} \\
	\end{tabular}
	\vspace{2pt}
	\hrule
\end{table}

\subsection*{Example 4. Data Analysis}
\label{subsec:real_data_analysis}

In this section, a cardiomyopathy microarray data \citep{redfern1999} from a transgenic mouse model of dilated cardiomyopathy, which has been studied by \citet{li2012}, is applied to investigate the performance of our coefficient. In this dataset, the response variable is the G protein-coupled receptor Ro1 expression level, and the predictors are the other 6319 related gene expression levels. Compared with its sample size \( n = 30 \), the dimension of predictors is enormous. Therefore, we aim to identify the influential genes for the expression level of Ro1 in mice.

\begin{figure}[htbp]
	\centering
	
	\begin{subfigure}[b]{0.45\textwidth}
		\centering
		\includegraphics[width=\linewidth]{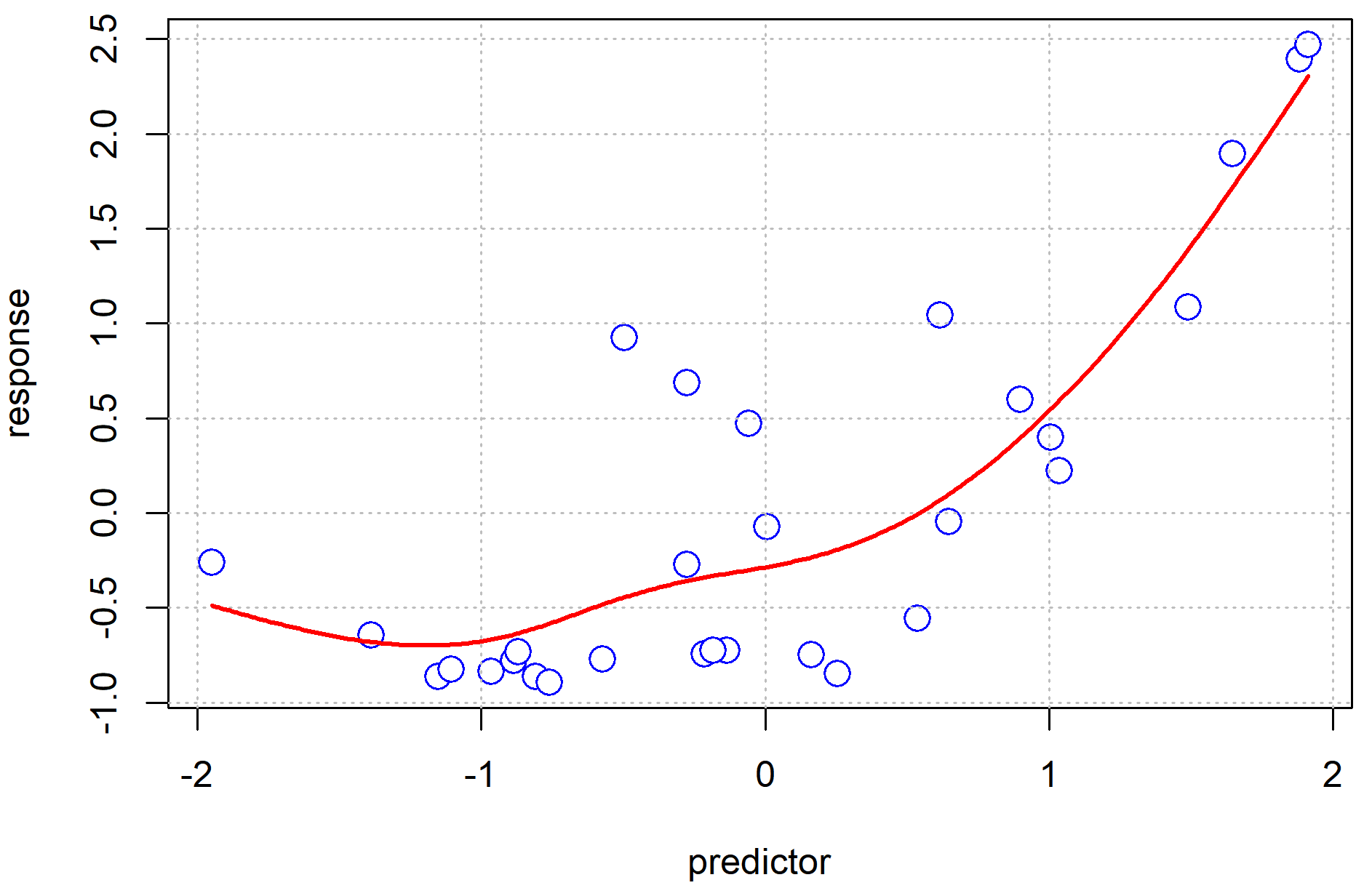}
		\caption{Msa.741.0}
		\label{subfig:msa741}
	\end{subfigure}
	\hfill
	\begin{subfigure}[b]{0.45\textwidth}
		\centering
		\includegraphics[width=\linewidth]{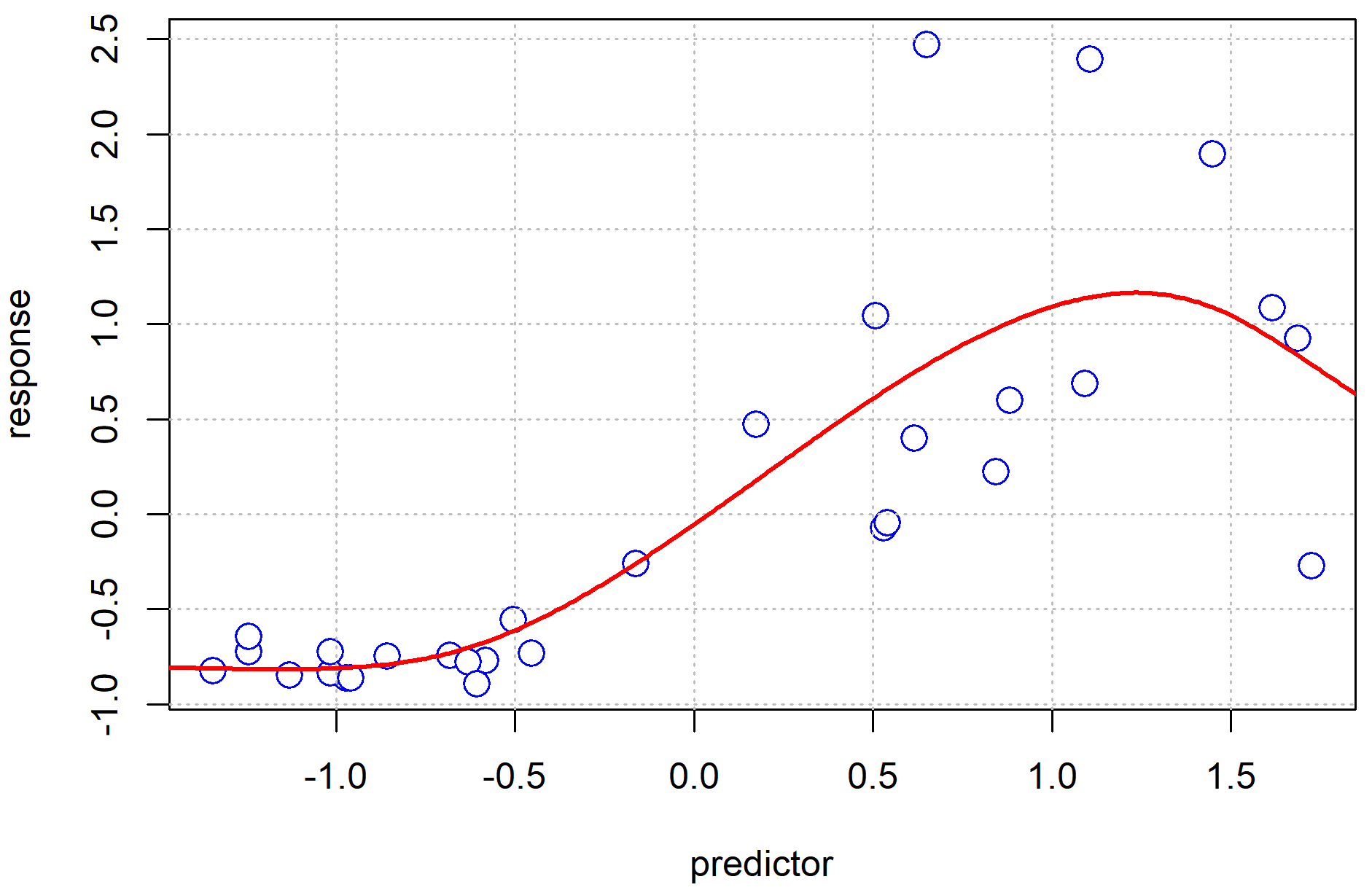}
		\caption{Msa.2134.0}
		\label{subfig:msa2134}
	\end{subfigure}
	
	\vspace{0.3cm}
	
	\caption{Scatter plots of the response variable against the top-ranked gene selected by DDC (left, Msa.741.0) and the top-ranked gene selected by the six competing measures (right, Msa.2134.0) in Example 4. 
		The fitted curves are obtained by generalized additive model to visualize the underlying dependence patterns.}
	\label{fig:top_genes}
\end{figure}

After standardizing the response variable and the predictors, we compute the coefficients between each predictor and the response variable. The most relevant 7 genes selected by DDC coefficients and their ranks in other coefficients are shown in Table~\ref{tab:gene_ranks}, which demonstrates that DDC and Chatterjee's coefficient exhibit higher concordance in the ranks of strongly correlated predictors. Moreover, Fig.~\ref{fig:top_genes} presents the scatter plots of the response variable versus the expression levels of gene Msa.741.0 and Msa.2134.0, the most correlated genes selected by DDC and the other six coefficients, respectively. To better indicate the trend of dependence, we fit the unknown link functions using the R \texttt{mgcv} package. The most correlated gene selected by DDC clearly achieves better performance with the adjusted \( R^2 \) of 73.1\% and the deviance explained of 76.4\%, in contrast to the adjusted \( R^2 \) of 65.9\% and the deviance explained of 70\% for the gene selected by the other coefficients. Here the deviance explained means the proportion of the null deviance explained by the proposed model, with a larger value indicating better performance. The results indicate that our proposed method tends to select predictors that exhibit a stronger association with the response variable.

\begin{table}[htbp]
	\centering
	\caption{Ranks of the seven top genes selected by DDC among all 6319 predictors in Example 4, together with their ranks under the six competing measures.}
	\label{tab:gene_ranks}
	\hrule
	\vspace{3pt}
	\begin{tabular}{c c c c c c c c}
		Method & Msa.741.0 & Msa.376.0 & Msa.2134.0 & Msa.2400.0 & Msa.21996.0 & Msa.3202.0 & Msa.1166.0 \\
		\cmidrule(lr){1-1} \cmidrule(lr){2-2} \cmidrule(lr){3-3} \cmidrule(lr){4-4} \cmidrule(lr){5-5} \cmidrule(lr){6-6} \cmidrule(lr){7-7} \cmidrule(lr){8-8}
		DDC & 1 & 2 & 3 & 4 & 5 & 6 & 7 \\
		Chatterjee & 6 & 3 & 1 & 60 & 8 & 10 & 5 \\
		DC & 10 & 116 & 1 & 7 & 72 & 108 & 6 \\
		PCor & 39 & 168 & 1 & 5 & 43 & 212 & 12 \\
		HSIC & 46 & 185 & 1 & 13 & 31 & 182 & 23 \\
		TauStar & 45 & 183 & 1 & 11 & 49 & 169 & 6 \\
		Ball & 21 & 182 & 1 & 14 & 53 & 90 & 3 \\
	\end{tabular}
	\vspace{3pt}
	\hrule
\end{table}

Then, we consider the dependent genes selected by the proposed method. For DDC and Chatterjee's coefficient, we directly calculate their \( p \)-values through their asymptotic normal distribution under the independent hypothesis. For the six permutation-based tests, 500 random permutations are used to estimate their \( p \)-values. The significance level of 0.05 is applied to detect the dependent predictors. Following the above process, the DDC coefficient identifies 930 dependent predictors.

\begin{figure}[htbp]
	\centering
	
	\begin{minipage}{0.32\textwidth}
		\centering
		\includegraphics[width=\linewidth]{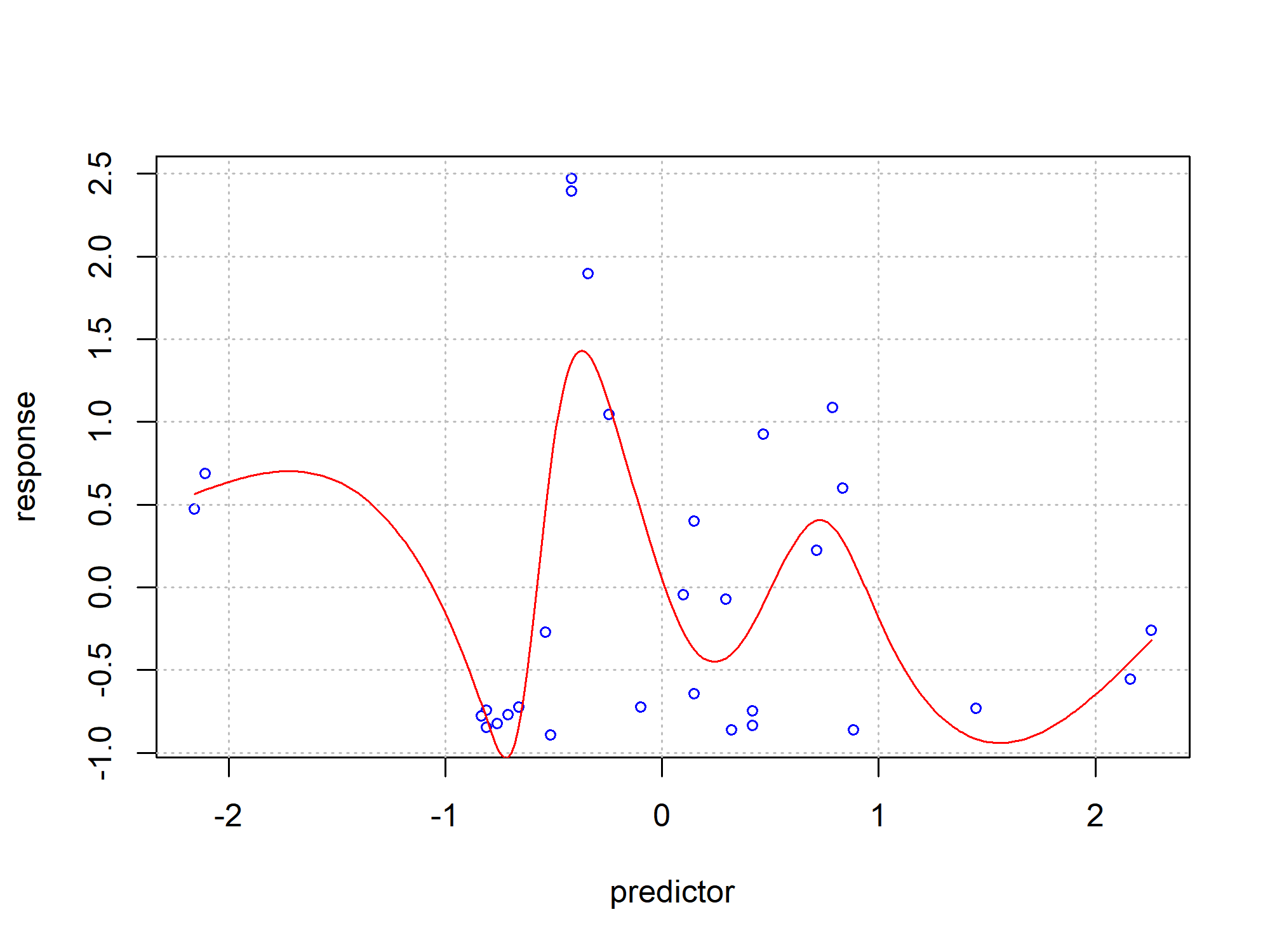}
		\caption*{Msa.763.0}
		\label{Msa.763.0}
	\end{minipage}
	\hfill
	\begin{minipage}{0.32\textwidth}
		\centering
		\includegraphics[width=\linewidth]{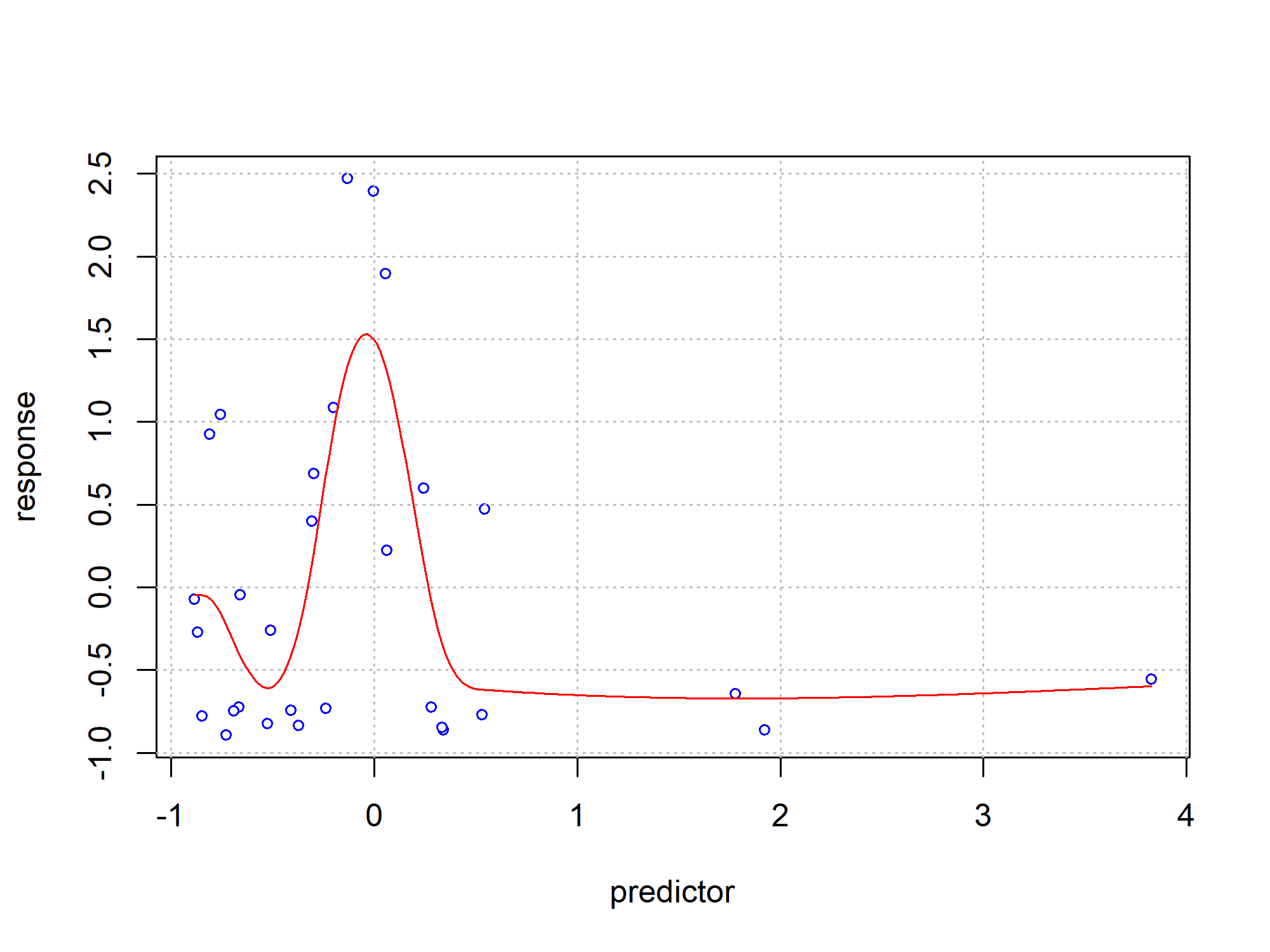}
		\caption*{Msa.119.0}
		\label{Msa.119.0}
	\end{minipage}
	\hfill
	\begin{minipage}{0.32\textwidth}
		\centering
		\includegraphics[width=\linewidth]{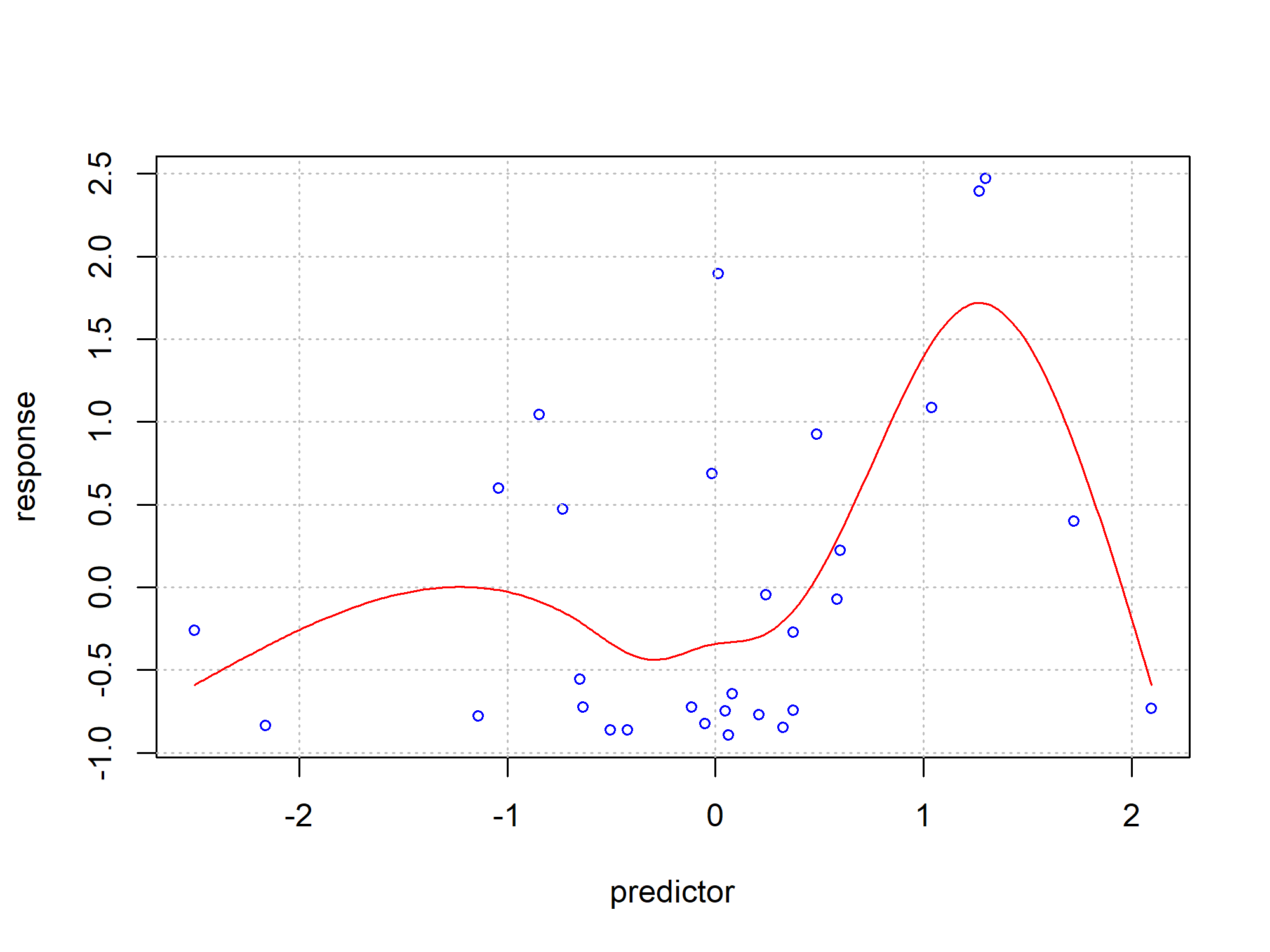}
		\caption*{Msa.20602.0}
		\label{Msa.20602.0}
	\end{minipage}
	
	\vspace{0.3cm} 
	
	\begin{minipage}{0.32\textwidth}
		\centering
		\includegraphics[width=\linewidth]{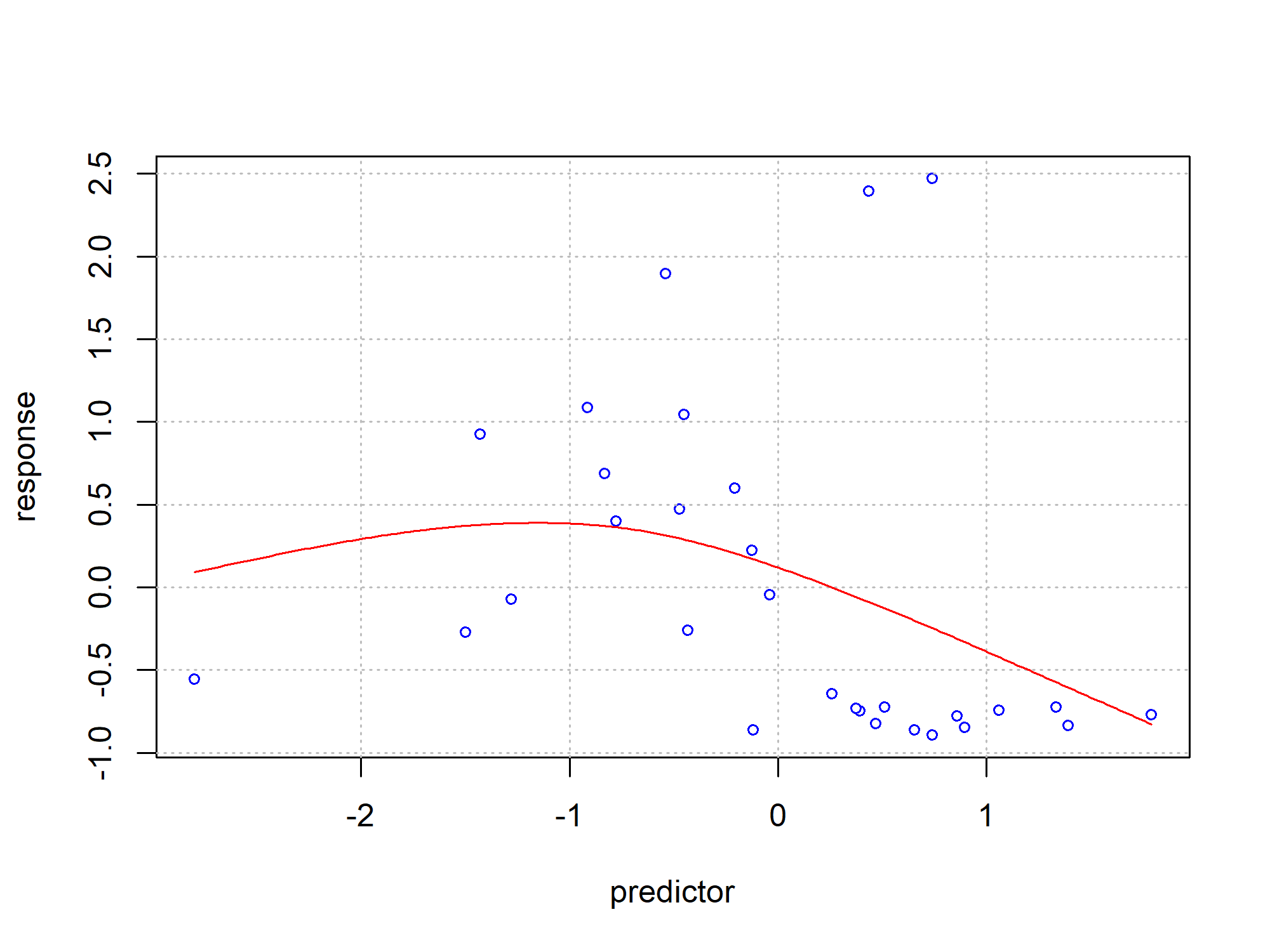}
		\caption*{Msa.13685.0}
		\label{Msa.13685.0}
	\end{minipage}
	\hfill
	\begin{minipage}{0.32\textwidth}
		\centering
		\includegraphics[width=\linewidth]{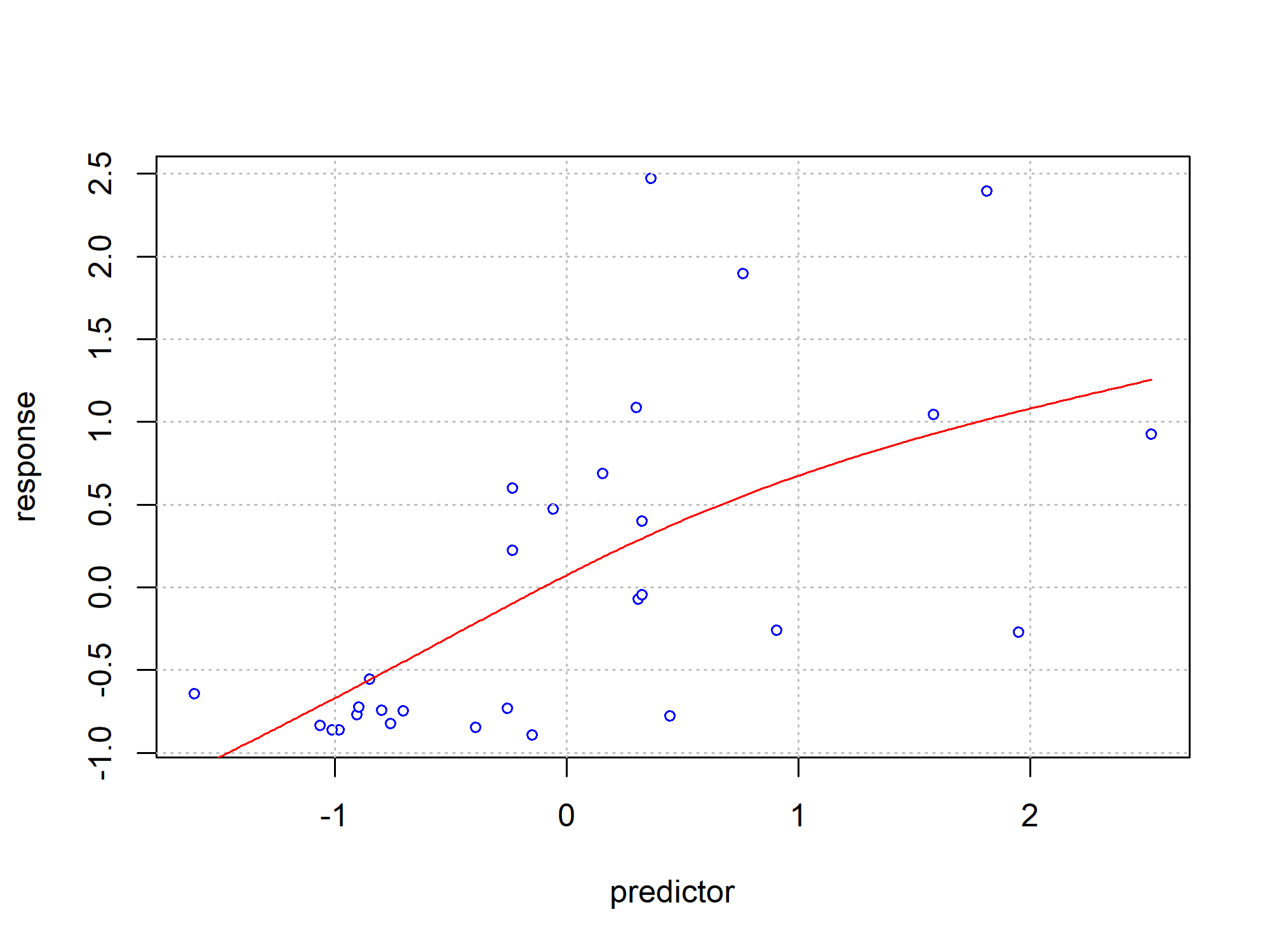}
		\caption*{Msa.3320.0}
		\label{Msa.3320.0}
	\end{minipage}
	\hfill
	\begin{minipage}{0.32\textwidth}
		\centering
		\includegraphics[width=\linewidth]{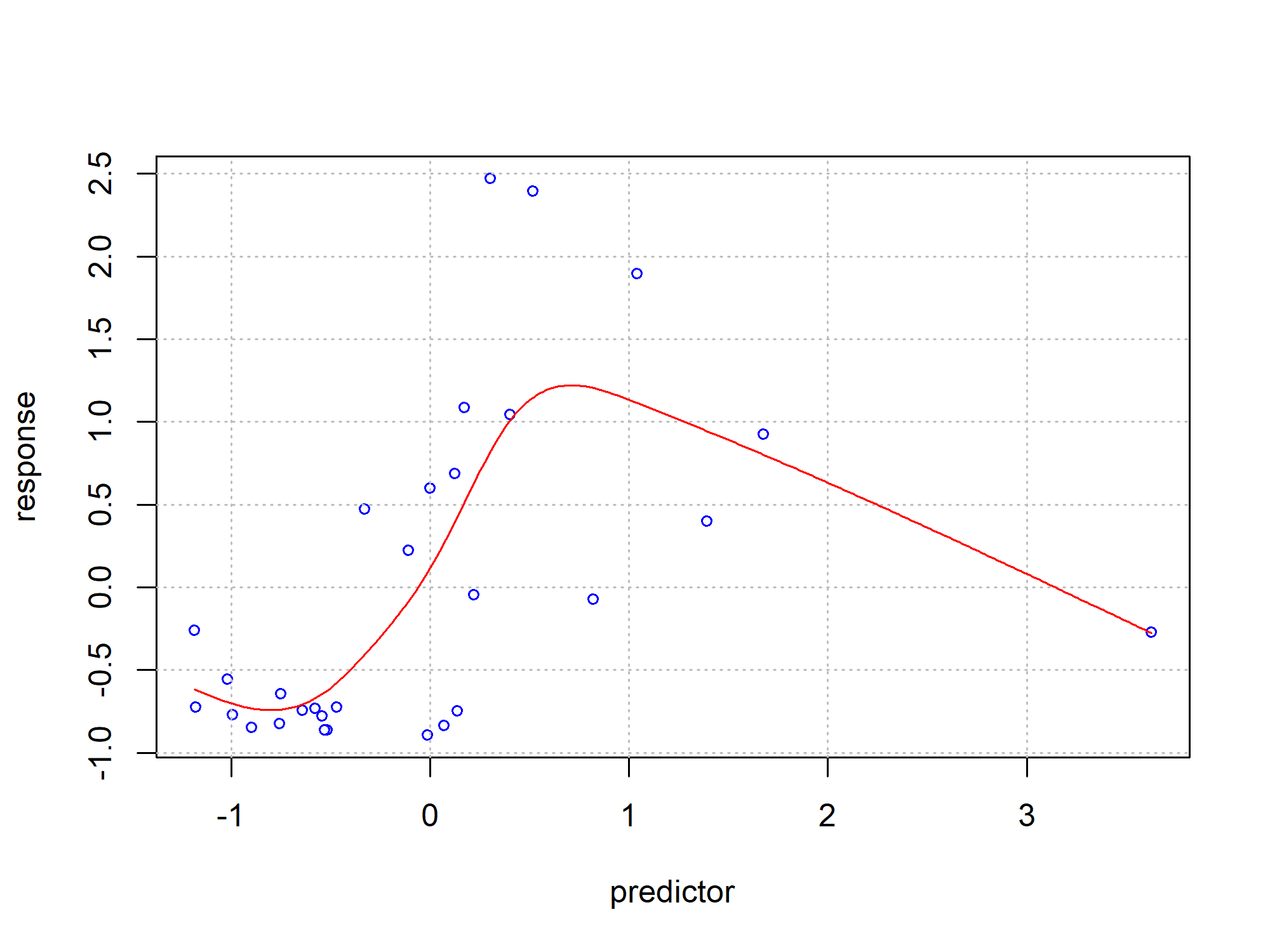}
		\caption*{Msa.42914.0}
		\label{Msa.42914.0}
	\end{minipage}
	
	\caption{Scatter plots with fitted curves of the response variable versus the three most significant genes exclusively identified by DDC (top row: Msa.763.0, Msa.119.0, and Msa.20602.0) and the three most significant genes jointly identified by all six permutation-based tests but not by DDC (bottom row: Msa.13685.0, Msa.3320.0, and Msa.42914.0). }
	\label{fig:ddc_vs_perm}
\end{figure}

We first compare DDC with the six permutation-based independence tests. The results show that 203 genes are uniquely identified by DDC but not by any other method, whereas 230 genes are detected by all of the six  permutation-based methods but not by DDC. Fig.~\ref{fig:ddc_vs_perm} presents the three most significant genes selected exclusively by DDC, Msa.763.0, Msa.119.0, and Msa.20602.0, with adjusted \( R^2 \) values of 0.435, 0.443, and 0.401, and deviance explained of 59.8\%, 56.3\%, and 51\%, respectively. For comparison, Fig.~\ref{fig:ddc_vs_perm} also displays the three most significant genes that are jointly selected by all six permutation-based tests but not by DDC: Msa.13685.0, Msa.3320.0, and Msa.42914.0. Their ranks are determined by the maximum \( p \)-value among the six tests. Notably, for each of these three genes, none of the permutation-generated statistics from any of the six methods exceeded the observed statistic. Their adjusted \( R^2 \) values are 0.121, 0.402, and 0.485, with deviance explained of 17.6\%, 43.3\%, and 56.1\%. In contrast, the genes identified exclusively by DDC exhibit stronger associations with the response. Moreover, as illustrated by the fitted curves in Fig.~\ref{fig:ddc_vs_perm}, the genes selected by DDC display more oscillatory patterns. These results suggest that DDC is more capable of detecting genes with stronger dependencies on the response variable, particularly when the underlying relationship exhibits obvious fluctuations.

\begin{figure}[htbp]
	\centering

\begin{minipage}{0.32\textwidth}
	\centering
	\includegraphics[width=\linewidth]{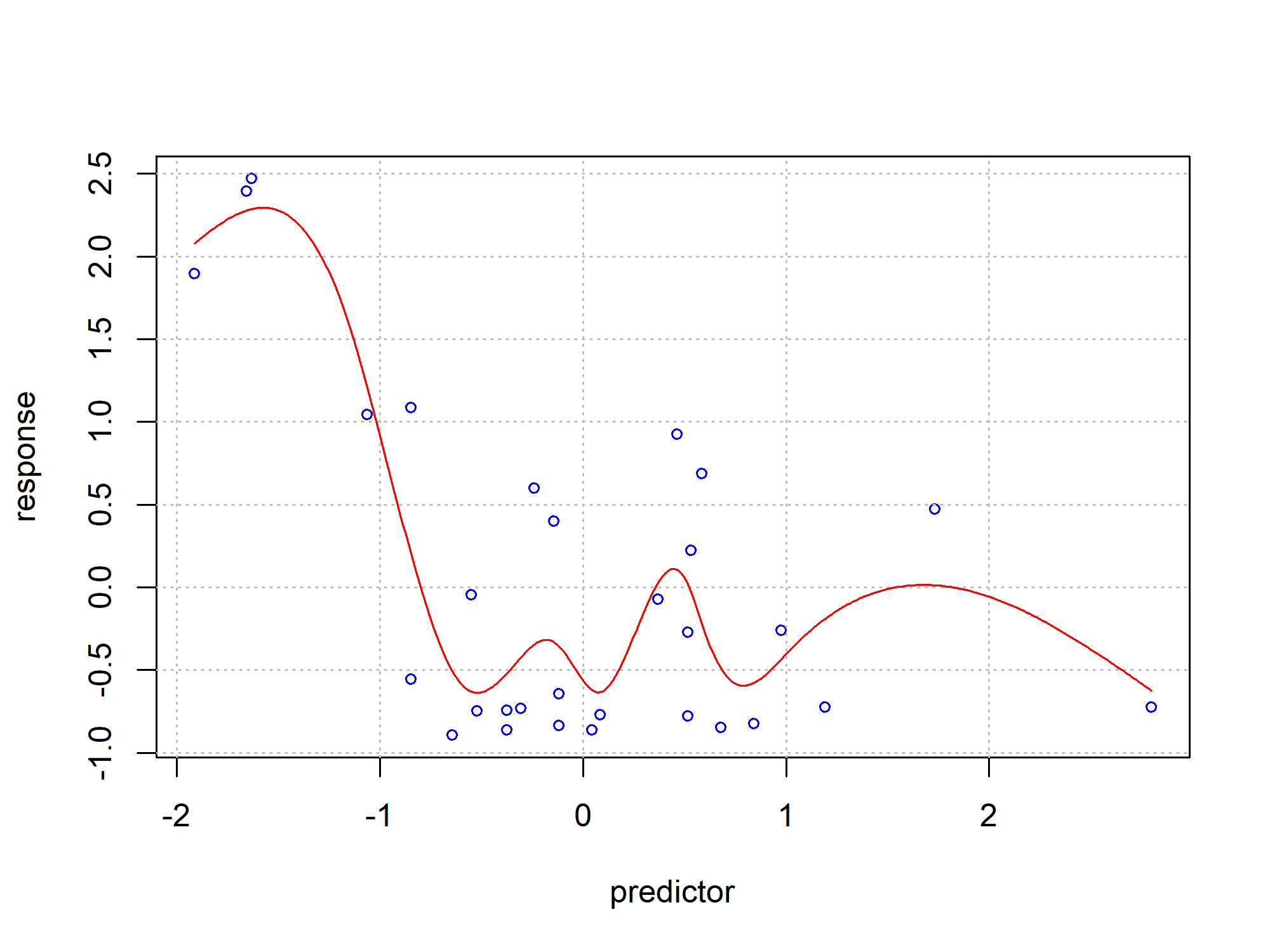}
	\caption*{Msa.6825.0}
	\label{Msa.6825.0}
\end{minipage}
\hfill
\begin{minipage}{0.32\textwidth}
	\centering
	\includegraphics[width=\linewidth]{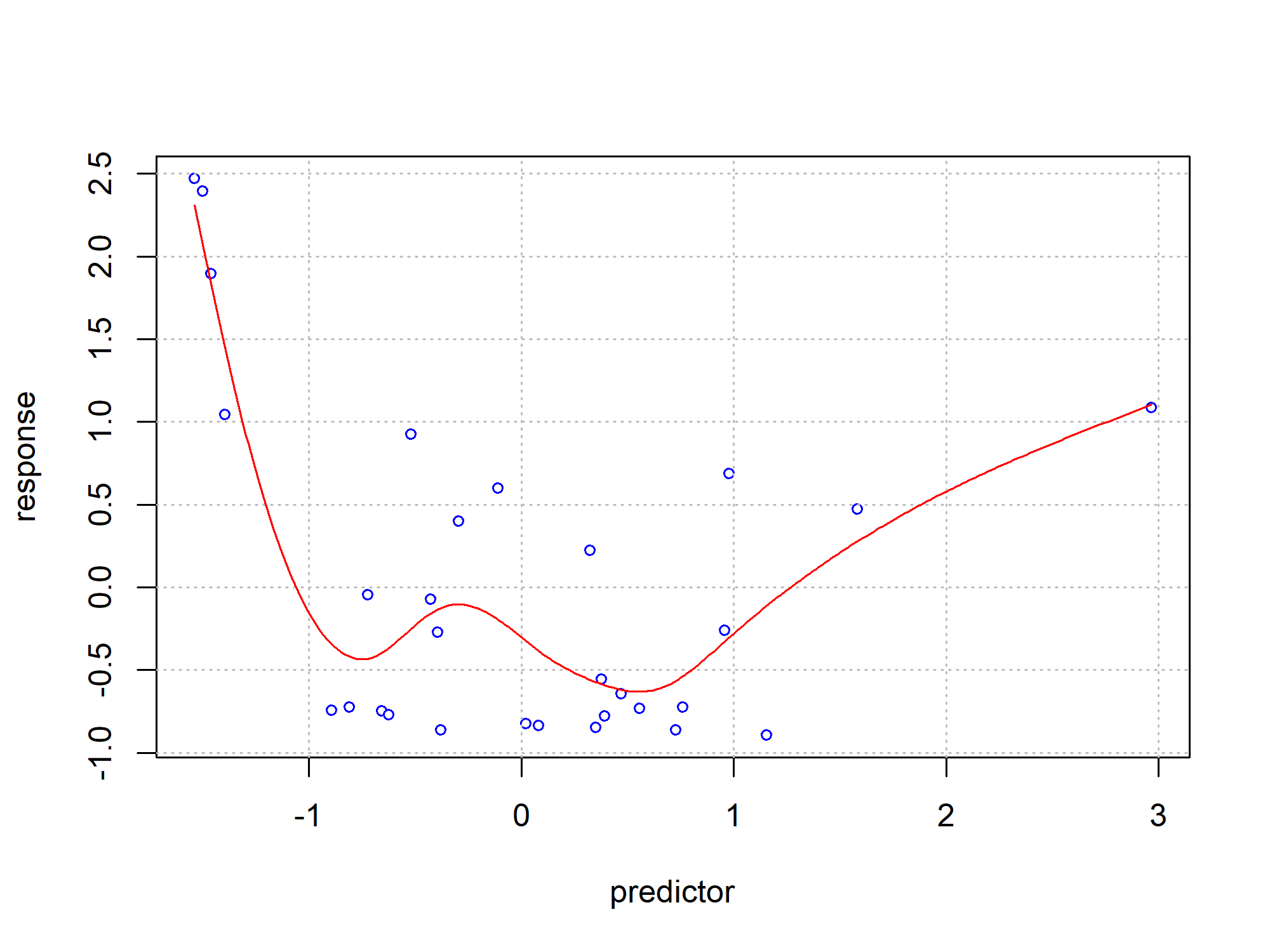}
	\caption*{Msa.28.0}
	\label{Msa.28.0}
\end{minipage}
\hfill
\begin{minipage}{0.32\textwidth}
	\centering
	\includegraphics[width=\linewidth]{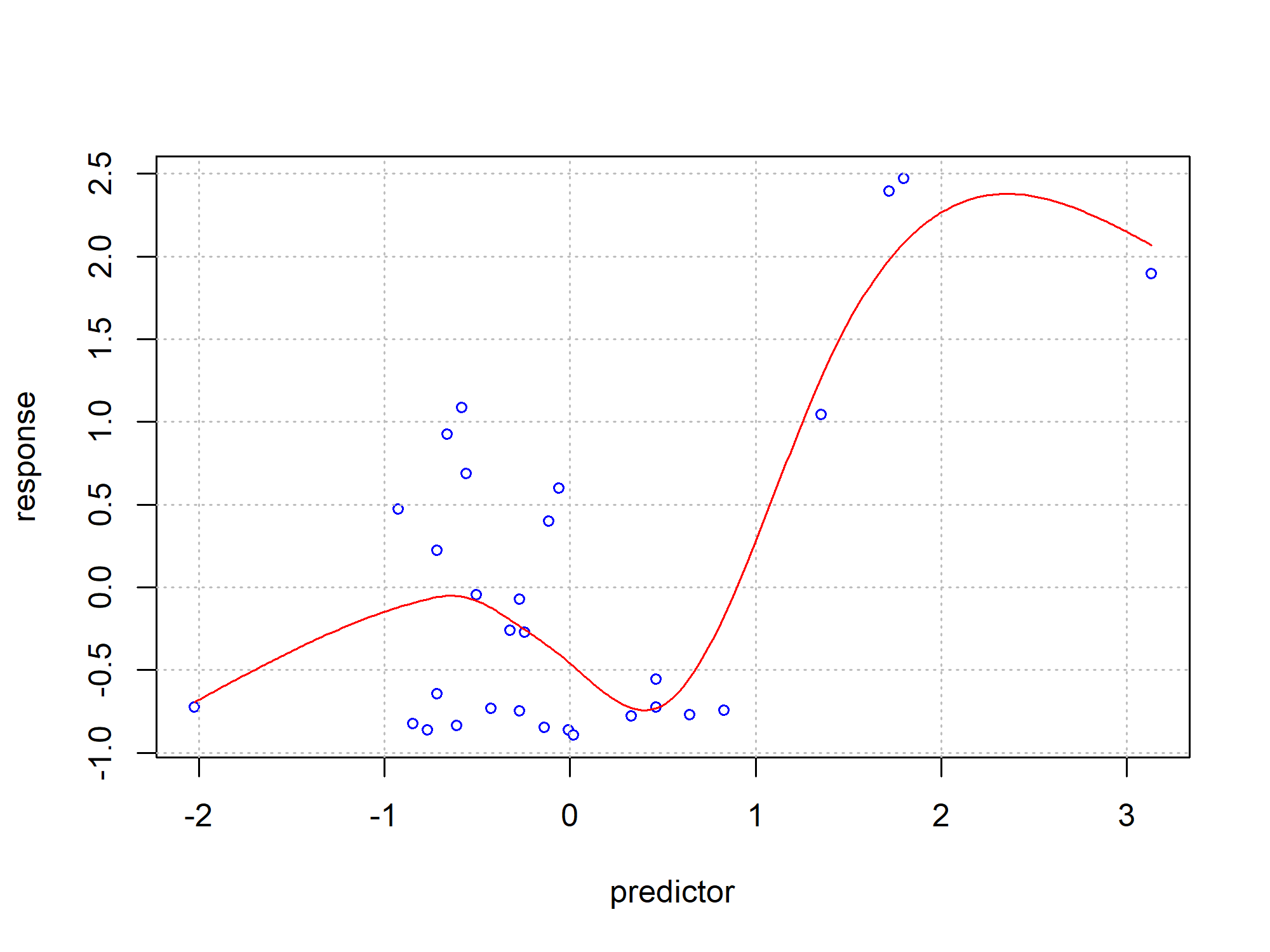}
	\caption*{Msa.28676.0}
	\label{Msa.28676.0}
\end{minipage}

\vspace{0.3cm} 

\begin{minipage}{0.32\textwidth}
	\centering
	\includegraphics[width=\linewidth]{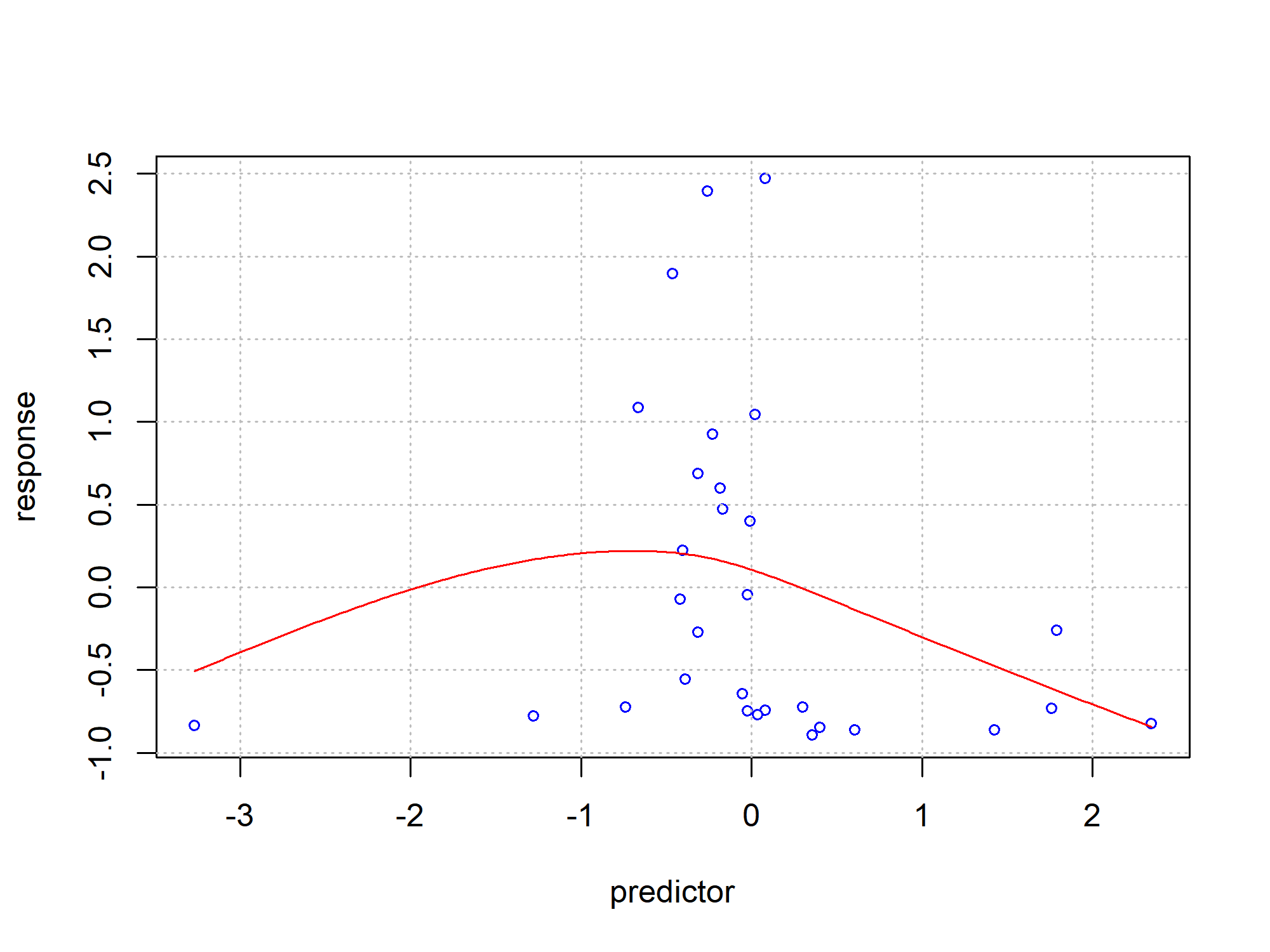}
	\caption*{Msa.19149.0}
	\label{Msa.19149.0}
\end{minipage}
\hfill
\begin{minipage}{0.32\textwidth}
	\centering
	\includegraphics[width=\linewidth]{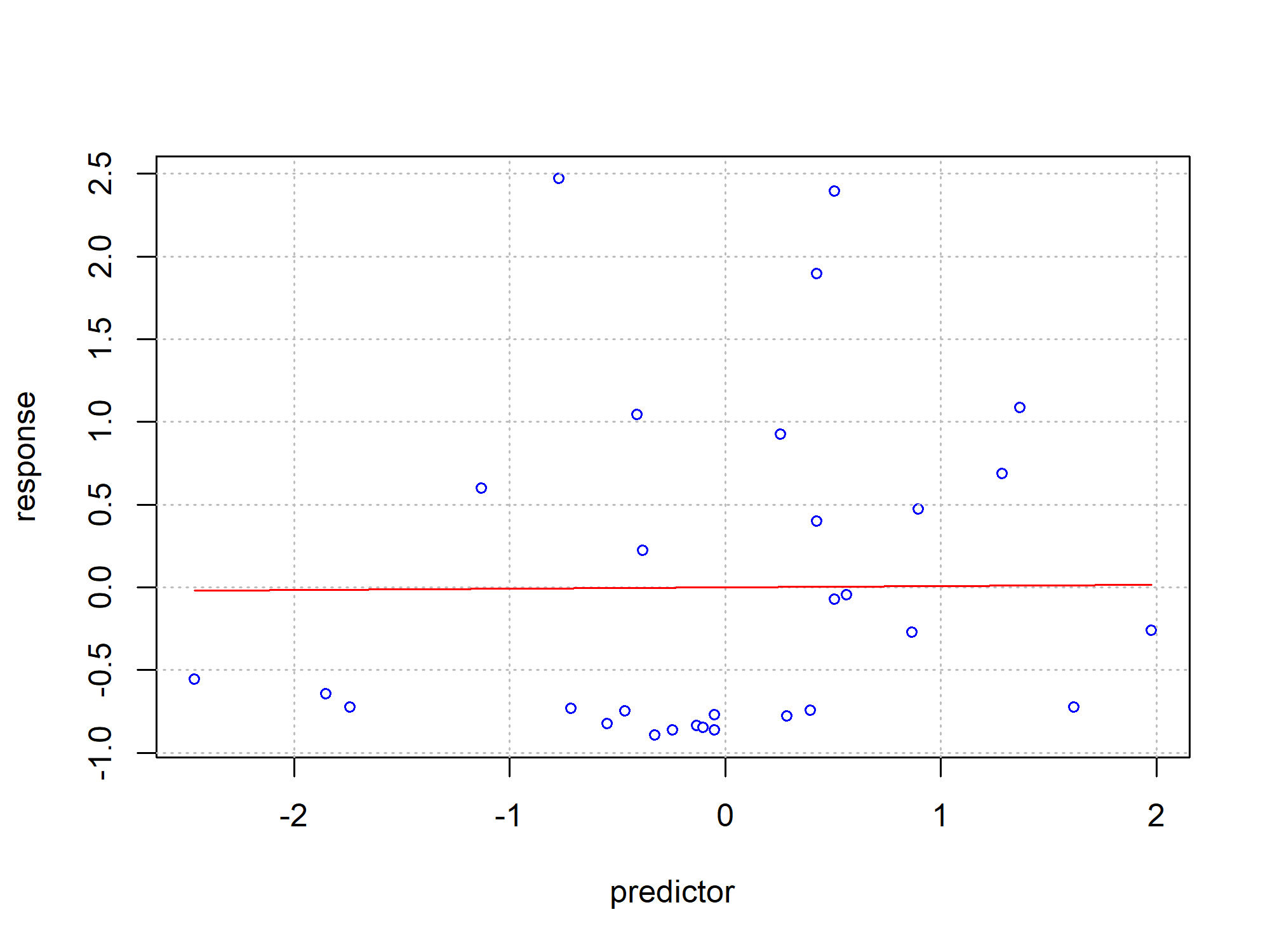}
	\caption*{Msa.4989.0}
	\label{Msa.4989.0}
\end{minipage}
\hfill
\begin{minipage}{0.32\textwidth}
	\centering
	\includegraphics[width=\linewidth]{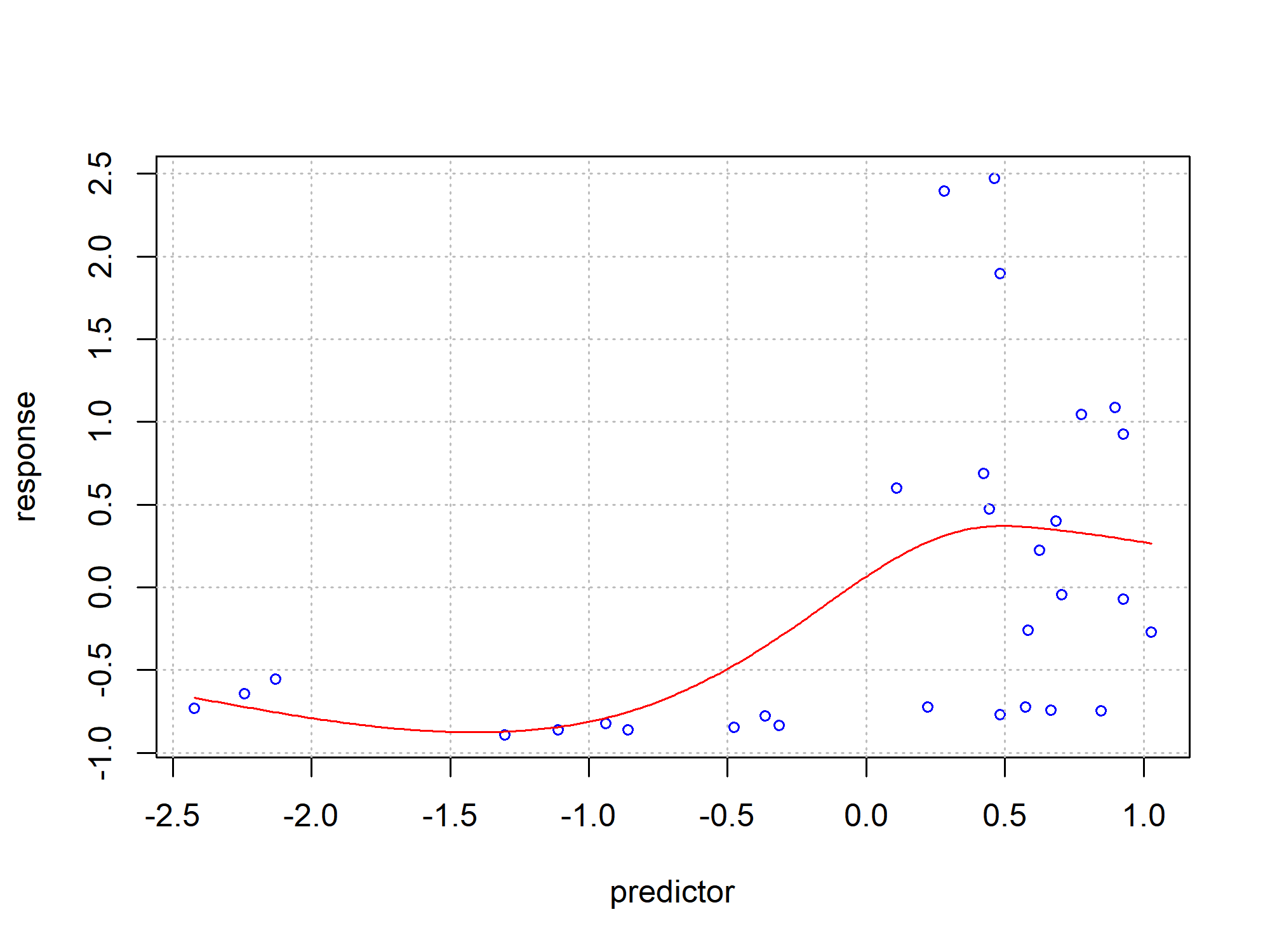}
	\caption*{Msa.10742.0}
	\label{Msa.10742.0}
\end{minipage}

\caption{Scatter plots with fitted curves of the response variable versus the three most significant genes exclusively identified by DDC but not by Chatterjee's coefficient (top row: Msa.6825.0, Msa.28.0, and Msa.28676.0) and the three most significant genes exclusively identified by Chatterjee's coefficient but not by DDC (bottom row: Msa.19149.0, Msa.4989.0, and Msa.10742.0).}
	\label{fig:ddc_vs_chatterjee}
\end{figure}

Next, we compare the performance of DDC with Chatterjee's coefficient. There are 397 genes exclusively identified by DDC but not by Chatterjee's coefficient. Among these genes, Msa.6825.0, Msa.28.0, and Msa.28676.0 are showing the smallest \( p \)-values, with adjusted \( R^2 \) values of 0.666, 0.709, and 0.625, and deviance explained of 75.7\%, 76.6\%, and 68.8\%, respectively. In contrast, 179 genes are detected as dependent by Chatterjee's coefficient but not by DDC, and among them, the three most significant genes Msa.19149.0, Msa.4989.0, and Msa.10742.0 show adjusted \( R^2 \) values of 0.090, 0.002, and 0.222, and deviance explained of 15.2\%, 0.302\%, and 29.8\%, respectively. The scatter plots of these aforementioned six genes are presented in Fig.~\ref{fig:ddc_vs_chatterjee}. The results demonstrate that the DDC-dependent but Chatterjee-independent predictors exhibit clearer functional dependence relationships with the response variable. In addition, the genes selected only by Chatterjee's coefficient show weaker fitting performance, and significant outliers in the response variable are observed in the dense cluster of predictors. This fact is potentially due to the fact that Chatterjee's coefficient neglects the metric differences in the response variable. Above all, compared to Chatterjee's coefficient, DDC is more effective in identifying genes that have a stronger functional association with the response variable.

\section{Discussion}\label{sec:discussion}

This study introduces a dependence measure \( DDC(\boldsymbol{X},Y) \) and proposes the differential distance correlation coefficient for estimating it. It is worth noting that \( DDC(\boldsymbol{X},Y) \) can also be applied to the setting where \( Y \) is categorical, as previously studied by \citet{dang2021}. However, their estimator has a complicated null distribution under independence and therefore requires permutation testing. In contrast, our estimation strategy offers a simpler solution: by converting the categorical variables to integer-valued variables in arbitrary ways, the proposed differential distance correlation can be applied directly, making independence testing more convenient.

For independence testing, it is important to note that the proposed coefficient is subject to the same limitation identified by \citet{shi2021} for Chatterjee's correlation. The fundamental reason is that both statistics rely primarily on right-nearest-neighbor differences after sorting along one coordinate. Under the two classes of local alternatives studied in \citet{shi2021}, such nearest-neighbor differences remain essentially unchanged, which results in limited testing power. Similar observations on the limitations of statistics based on the adjacent neighbour information have also been discussed in \citet{cao2020} and \citet{bickel2022}. To address this issue, we can adopt the approach of \citet{lin2022}, using multiple right-nearest-neighbors' differences to estimate \( DDC(\boldsymbol{X},Y) \), thereby increasing the testing power.

Regarding the moment limitations in \( DDC(\boldsymbol{X},Y) \), in addition to the alternative approach discussed in Section~\ref{subsec:extension}, further strategies can be adopted to address this limitation. One viable method is modifying the weight function \( w(\|\boldsymbol{t}\|) \) in \eqref{eq:simplified_integral} to express the independence measure in a form that is not subject to the moment limitations. For example, by defining the weight function as the Gaussian kernel function \( w_1(\boldsymbol{t}) = \exp(-\|\boldsymbol{t}\|^2 / 2) \), the Cauchy kernel function \( w_2(\boldsymbol{t}) = 1/(1 + \|\boldsymbol{t}\|^2) \), and the Laplacian kernel function \( w_3(\boldsymbol{t}) = \exp(-\|\boldsymbol{t}\|) \), the term \( \|\boldsymbol{X}_1 - \boldsymbol{X}_2\| \) in \( DDC(\boldsymbol{X},Y) \) would be correspondingly replaced by \( 1 - \exp(-\|\boldsymbol{X}_1 - \boldsymbol{X}_2\|^2 / 2) \), \( 1 - \exp(-\|\boldsymbol{X}_1 - \boldsymbol{X}_2\|) \), and \( 1 - 1/(1 + \|\boldsymbol{X}_1 - \boldsymbol{X}_2\|^2) \). These substitutions can be easily derived through the Fourier transformation. However, it should be noted that when different weight functions are employed, the computational cost of estimating the result formulations may not be reduced.

Moreover, the proposed coefficient is restricted to settings where \( Y \) is univariate. A promising direction for future work is to replace the right-nearest-neighbor estimator with the nearest-neighbor, which would allow the definition of \( Y \) to be extended naturally to the multivariate case. However, such a modification would alter the structure of the estimator, and its convergence properties and asymptotic behavior require systematic and rigorous theoretical investigation, which constitutes an important direction for future research.

\section{Conclusion}\label{sec:conclusion}

In this article, we propose a simple coefficient that consistently estimates an interpretable measure of dependence, which is 0 if and only if the random vector \( \boldsymbol{X} \) and the variable \( Y \) are independent and equal to 1 if and only if \( Y \) determines \( \boldsymbol{X} \) almost surely. We further establish its asymptotic normality under the independence hypothesis and provide an alternative method that avoids moment limitations. Simulation experiments demonstrate that the proposed method offers substantial computational savings compared with several permutation-based independence tests, while being more effective at detecting oscillatory dependence relationships in both univariate and multivariate response models. Moreover, in an application to a microarray dataset, the proposed coefficient identifies predictors exhibiting stronger dependence on the response than competing approaches. Finally, several limitations of our work and directions for further research are proposed in the discussion section.

\section*{Acknowledgment}

The authors are grateful to the reviewers for their helpful comments and constructive suggestions, which have significantly improved the quality of this paper. This research is supported by the funds of the National Natural Science Foundation of China (62171018).

\section*{Appendix}
\label{sec:appendix}
The Appendix provides the technical proofs of Theorem~\ref{thm:ddc_properties}, Theorem~\ref{thm:asymptotic_normality}, the two cases in Section~\ref{subsec:asymptotic_normality}, and Proposition~\ref{prop:extension}.

\begin{proof}[\normalfont\bfseries Proof of Theorem~\ref{thm:ddc_properties}: Some properties of \( DDC(\boldsymbol{X},Y) \) and \( DDC_n(\boldsymbol{X},Y) \).]
	
\mbox{}\\[0.1em]  
	
\noindent{(i)} First, a decomposition formula is presented in Lemma~1 for the following proof.

\begin{lemma}
	\label{lem:decomposition}
	For random vector \( \boldsymbol{X} \in \mathbb{R}^p \), arbitrary vector \( \boldsymbol{t} \in \mathbb{R}^p \), and random variable \( Y \in \mathbb{R} \),
	\[
	E \left[ \left| e^{i\langle \boldsymbol{t},\boldsymbol{X} \rangle} - E \left( e^{i\langle \boldsymbol{t},\boldsymbol{X} \rangle} \right) \right|^2 \right] = E \left[ \left| e^{i\langle \boldsymbol{t},\boldsymbol{X} \rangle} - E \left( e^{i\langle \boldsymbol{t},\boldsymbol{X} \rangle} \mid Y \right) \right|^2 \right] + E \left[ \left| E \left( e^{i\langle \boldsymbol{t},\boldsymbol{X} \rangle} \mid Y \right) - E \left( e^{i\langle \boldsymbol{t},\boldsymbol{X} \rangle} \right) \right|^2 \right].
	\]
\end{lemma}

\begin{proof}
	\begin{align*}
		&E \left[ \left| e^{i\langle \boldsymbol{t},\boldsymbol{X} \rangle} - E \left( e^{i\langle \boldsymbol{t},\boldsymbol{X} \rangle} \right) \right|^2 \right] = 1 - \left| E \left( e^{i\langle \boldsymbol{t},\boldsymbol{X} \rangle} \right) \right|^2 \\
		&= 1 - E \left[ \left| E \left( e^{i\langle \boldsymbol{t},\boldsymbol{X} \rangle} \mid Y \right) \right|^2 \right] + E \left[ \left| E \left( e^{i\langle \boldsymbol{t},\boldsymbol{X} \rangle} \mid Y \right) \right|^2 \right] - \left| E \left( e^{i\langle \boldsymbol{t},\boldsymbol{X} \rangle} \right) \right|^2 \quad \\
		&= E \left[ \left| e^{i\langle \boldsymbol{t},\boldsymbol{X} \rangle} - E \left( e^{i\langle \boldsymbol{t},\boldsymbol{X} \rangle} \mid Y \right) \right|^2 \right] + E \left[ \left| E \left( e^{i\langle \boldsymbol{t},\boldsymbol{X} \rangle} \mid Y \right) - E \left( e^{i\langle \boldsymbol{t},\boldsymbol{X} \rangle} \right) \right|^2 \right].
	\end{align*}
\end{proof}

It is easy to obtain that
\[
\int_{\mathbb{R}^p} \left( 1 - \left| E \left( e^{i\langle \boldsymbol{t},\boldsymbol{X} \rangle} \right) \right|^2 \right) w(\boldsymbol{t}) d\boldsymbol{t} = E \left[ \left\| \boldsymbol{X}_1 - \boldsymbol{X}_2 \right\| \right].
\]

Therefore, the range of \( DDC(\boldsymbol{X},Y) \) is [0,1], and it is natural \( DDC(\boldsymbol{X},Y) = 0 \) if and only if \( \boldsymbol{X} \) and \( Y \) are independent. Next, we prove the functional dependence property. If \( DDC(\boldsymbol{X},Y) = 1 \), for all \( \boldsymbol{t} \in \mathbb{R}^p \), there will be \( e^{i\langle \boldsymbol{t},\boldsymbol{X} \rangle} = E(e^{i\langle \boldsymbol{t},\boldsymbol{X} \rangle} \mid Y) \), which indicates that \(\sin \langle \boldsymbol{t}, \boldsymbol{X} \rangle = E(\sin \langle \boldsymbol{t}, \boldsymbol{X} \rangle \mid Y)\) almost surely. As \( E[\| \boldsymbol{X} \|] < +\infty \), according to the Lebesgue's dominated convergence theorem (denoted as DCT in the following), there will be
\[
\boldsymbol{X} \cos \langle \boldsymbol{t}, \boldsymbol{X} \rangle = \nabla_{\boldsymbol{t}} \sin \langle \boldsymbol{t}, \boldsymbol{X} \rangle = \nabla_{\boldsymbol{t}} E (\sin \langle \boldsymbol{t}, \boldsymbol{X} \rangle \mid Y) = E \left( \boldsymbol{X} \cos \langle \boldsymbol{t}, \boldsymbol{X} \rangle \mid Y \right) 
\]
almost surely, where \( \nabla_{\boldsymbol{t}} \) is the symbol of gradient. Let \( \boldsymbol{t} \to \boldsymbol{0} \), \( \boldsymbol{X} = E(\boldsymbol{X} \mid Y) \) almost surely according to the DCT. Since \( E(\boldsymbol{X} \mid Y) \) is a measurable function vector of \( Y \), this property is proved.\\
\\
\noindent{(ii)} By the definitions of \( DDC(\boldsymbol{X},Y) \) and \( DDC_n(\boldsymbol{X},Y) \), this property can be easily proved.\\
\\
\noindent{(iii)} Firstly, we rearrange \( \{(\boldsymbol{X}_i, Y_i)\}_{i=1}^n \) as \( \{(\boldsymbol{X}_{(i)}, Y_{(i)})\}_{i=1}^n \) such that \( Y_{(1)} \leq ... \leq Y_{(n)} \), and define \( Q_n \) and \( Q \) as below:
\[
Q_n = \frac{1}{n-1} \sum_{i=1}^{n-1} \left\| \boldsymbol{X}_{(i)} - \boldsymbol{X}_{(i+1)} \right\|, \quad Q = E_{Y} \left[ E \left( \left\| \boldsymbol{X}_1 - \boldsymbol{X}_2 \right\| \mid Y_1 = Y_2 = Y \right) \right].
\]
Next, we prove the following lemma.

\begin{lemma}
	\label{lem:Qn_convergence}
	\( Q_n \to Q \) almost surely as \( n \to \infty \).
\end{lemma}

\begin{proof}
	Define
	\begin{equation*}
		\label{eq:main_def}
		\begin{aligned}
			\zeta_n(\boldsymbol{t})
			&= 1 - \frac{1}{n-1} \sum_{i=1}^{n-1}
			\cos\langle \boldsymbol{t}, \boldsymbol{X}_{(i)} - \boldsymbol{X}_{(i+1)} \rangle,  ~
			Q_{n,\delta}
			= \int_{\mathbb{R}^p \setminus B_\delta}
			\zeta_n(\boldsymbol{t}) w(\boldsymbol{t}) \, d\boldsymbol{t},\\
			Q_\delta 
			&= \int_{\mathbb{R}^p \setminus B_\delta}
			\left(
			1 - E\!\left[
			\left|
			E\!\left(
			e^{i\langle \boldsymbol{t}, \boldsymbol{X} \rangle}
			\mid Y
			\right)
			\right|^2
			\right]
			\right)
			w(\boldsymbol{t}) \, d\boldsymbol{t},
		\end{aligned}
	\end{equation*}
	
\noindent	where \( B_\delta = \{\boldsymbol{t} \in \mathbb{R}^p : \|\boldsymbol{t}\| < \delta\} \). \( Q_{n,\delta} \) is bounded and satisfies
	\begin{equation}
		\label{eq:Qn_bound}
		Q_{n,\delta} \leq 2 \int_{\mathbb{R}^p\setminus B_\delta} w(\boldsymbol{t}) d\boldsymbol{t} = 2 \frac{\Gamma\left(\frac{p+1}{2}\right)}{\sqrt{\pi} \Gamma\left(\frac{p}{2}+1\right)} \delta^{-p}= C(p, \delta) ,
	\end{equation}
	where \( C(p, \delta) \) is defined in the obvious manner. The proof is divided into the following three parts:
	\begin{enumerate}[label=\alph*.]
		\item \(\displaystyle \limsup_{\delta \to 0} \limsup_{n \to \infty} |Q_{n,\delta} - Q_n| = 0\) a.s.;
		\item \( Q_{n,\delta} \to Q_\delta \) as \( n \to \infty \) a.s.;
		\item \(\displaystyle \lim_{\delta \to 0} Q_\delta = Q\),
	\end{enumerate}
	where a.s. is the abbreviation of almost surely.
	
	\textit{Part 1}. For \( \boldsymbol{t} = (t_1, \dots, t_p) \in \mathbb{R}^p \), define the function
	\begin{equation}
		\label{eq:G_delta_def}
		G(\delta) = \int_{B_\delta} (1 - \cos t_1) w(\boldsymbol{t}) d\boldsymbol{t}.
	\end{equation}
	It is easy to demonstrate that \( G(\delta) \leq 1 \) and \(\displaystyle \lim_{\delta \to 0} G(\delta) = 0\). Next, \( |Q_{n,\delta} - Q_n| \) is discussed with fixed \( n \) and \( \delta \) as follows:
	\begin{align*}
		|Q_{n,\delta} - Q_n| &= \left| \int_{B_\delta} \zeta_n(\boldsymbol{t}) w(\boldsymbol{t}) d\boldsymbol{t} \right| 
		= \frac{1}{n-1} \sum_{i=1}^{n-1} \int_{B_\delta} \left(1 - \cos\langle \boldsymbol{t}, \boldsymbol{X}_{(i)} - \boldsymbol{X}_{(i+1)} \rangle\right) w(\boldsymbol{t}) d\boldsymbol{t} \\
		&= \frac{1}{n-1} \sum_{i=1}^{n-1} \|\boldsymbol{X}_{(i)} - \boldsymbol{X}_{(i+1)}\| \, G\!\left(\delta\|\boldsymbol{X}_{(i)} - \boldsymbol{X}_{(i+1)}\|\right) \\
		&\leq \frac{1}{n-1} \sum_{i=1}^{n-1} 2 \max\left\{ \|\boldsymbol{X}_{(i)}\|, \|\boldsymbol{X}_{(i+1)}\| \right\} G\!\left(2\delta\max\left\{ \|\boldsymbol{X}_{(i)}\|, \|\boldsymbol{X}_{(i+1)}\| \right\}\right) \\
		&\leq \frac{2}{n-1} \sum_{i=1}^{n-1} \left( \|\boldsymbol{X}_{(i)}\| G\!\left(2\delta \|\boldsymbol{X}_{(i)}\|\right) + \|\boldsymbol{X}_{(i+1)}\| G\!\left(2\delta \|\boldsymbol{X}_{(i+1)}\|\right) \right) 
		\leq \frac{4}{n-1} \sum_{i=1}^{n} \|\boldsymbol{X}_i\| G\!\left(2\delta \|\boldsymbol{X}_i\|\right).
	\end{align*}
	By the strong law of large numbers,
	\[
	\limsup_{n \to \infty} |Q_{n,\delta} - Q_n| \leq 4E\left[ \|\boldsymbol{X}\| G\!\left(2\delta \|\boldsymbol{X}\|\right) \right]
	\]
	almost surely. Therefore, by Fatou's lemma,
	\[
	\limsup_{\delta \to 0} \limsup_{n \to \infty} |Q_{n,\delta} - Q_n| = 0
	\]
	almost surely.
	
	\textit{Part 2.} Denote \( Q_{n,\delta} \) as \( H_\delta(\boldsymbol{X}_1, \dots, \boldsymbol{X}_n) \), and generate an independent sample \( \tilde{\boldsymbol{X}} \) from \( \boldsymbol{X} \), we derive the following result:
	\begin{align*}
&|H_\delta(\boldsymbol{X}_1, \dots, \tilde{\boldsymbol{X}}, \dots, \boldsymbol{X}_n) - H_\delta(\boldsymbol{X}_1, \dots, \boldsymbol{X}_i, \dots, \boldsymbol{X}_n)| \\ 
&= \frac{1}{n-1} \left| \int_{\mathbb{R}^p\setminus B_\delta} \left[ \cos\langle \boldsymbol{t}, \boldsymbol{X}_{(j-1)} - \boldsymbol{X}_{(j)} \rangle + \cos\langle \boldsymbol{t}, \boldsymbol{X}_{(j)} - \boldsymbol{X}_{(j+1)} \rangle - \cos\langle \boldsymbol{t}, \boldsymbol{X}_{(j-1)} - \tilde{\boldsymbol{X}} \rangle \right. \right.\\
& \quad \left. \left. - \cos\langle \boldsymbol{t}, \tilde{\boldsymbol{X}} - \boldsymbol{X}_{(j+1)} \rangle \right] w(\boldsymbol{t}) d\boldsymbol{t} \right|  
\leq \frac{4}{n-1} \int_{\mathbb{R}^p\setminus B_\delta} w(\boldsymbol{t}) d\boldsymbol{t} = \frac{2}{n-1} C(p, \delta),
	\end{align*}
	where \( j = \omega(i) \) and \( j \notin \{1, n\} \). Here \( \omega(i) \) denotes the rank of \( Y_i \). When \( j \in \{1, n\} \), the result is less than or equal to \( C(p, \delta)/(n-1) \). According to McDiarmid's inequality \citep{zhang2021a}, for any \( n \) and \( \varepsilon > 0 \), we have
	\begin{equation}
		\label{eq:McDiarmid}
		P\left(|Q_{n,\delta} - E[Q_{n,\delta}]| \geq \varepsilon\right) \leq 2 \exp\left( -\frac{2(n-1)^2 \varepsilon^2}{(4n-6)C^2(p,\delta)} \right) \leq 2 \exp\left( -\frac{(n-1)\varepsilon^2}{2C^2(p,\delta)} \right).
	\end{equation}
	
	Next, let's demonstrate \(\displaystyle \lim_{n \to \infty} E[Q_{n,\delta}] = Q_\delta\). Define \( f_c(Y) = E[\cos\langle \boldsymbol{t}, \boldsymbol{X} \rangle \mid Y] \), \( f_s(Y) = E[\sin\langle \boldsymbol{t}, \boldsymbol{X} \rangle \mid Y] \), and \( N(i) \) as follows:
	\[
	N(i) = 
	\begin{cases}
		\omega^{-1}(\omega(i)+1), & \text{if } \omega(i) < n; \\
		i, & \text{if } \omega(i) = n.
	\end{cases}
	\]
	Then \( \zeta_n(\boldsymbol{t}) \), \( E[Q_{n,\delta}] \) and \( Q_\delta \) can be expressed as
	\begin{align*}
		\zeta_n(\boldsymbol{t}) &= 1 - \frac{1}{n-1} \sum_{i=1}^{n-1} \cos\langle \boldsymbol{t}, \boldsymbol{X}_{(i)} - \boldsymbol{X}_{(i+1)} \rangle = \frac{n}{n-1} \left( 1 - \frac{1}{n} \sum_{i=1}^{n} \cos\langle \boldsymbol{t}, \boldsymbol{X}_i - \boldsymbol{X}_{N(i)} \rangle \right), \\
		E[Q_{n,\delta}] &= \int_{\mathbb{R}^p\setminus B_\delta} \frac{n}{n-1} \left( 1 - E \left[ f_c(Y_1) f_c(Y_{N(1)}) + f_s(Y_1) f_s(Y_{N(1)}) \right] \right) w(\boldsymbol{t}) d\boldsymbol{t}, \\
		Q_\delta &= \int_{\mathbb{R}^p\setminus B_\delta} \left( 1 - E \left[ f_c(Y)^2 + f_s(Y)^2 \right] \right) w(\boldsymbol{t}) d\boldsymbol{t}.
	\end{align*}
	From Corollary 9.9 in \citet{chatterjee2021}, for measurable function \( f : \mathbb{R} \to \mathbb{R} \), \( f(Y_1) - f(Y_{N(1)}) \to 0 \) in probability as \( n \to \infty \). Therefore, as \( f_c \) and \( f_s \) are bounded, by the dominated convergence theorem, for all \( \delta > 0 \),
	\begin{equation}
		\label{eq:limit_Qndelta}
		\lim_{n \to \infty} E[Q_{n,\delta}] = \int_{\mathbb{R}^p\setminus B_\delta} \left( 1 - E \left[ f_c(Y_1)^2 + f_s(Y_1)^2 \right] \right) w(\boldsymbol{t}) d\boldsymbol{t} = Q_\delta.
	\end{equation}
	Combining with the McDiarmid inequality result, we obtain that \( Q_{n,\delta} \to Q_\delta \) almost surely for all \( \delta > 0 \).
	
	\textit{Part 3.} By the monotone convergence theorem, \( Q_\delta \to Q \) as \( \delta \to 0 \).
	
	Above all, \( Q_n \to Q \) almost surely.
\end{proof}

According to the properties of U-statistics \citep{serfling1980}, $\hat{\Delta}_n \to \Delta$ almost surely as $n \to \infty$. As $\boldsymbol{X}$ is non-degenerate, $E[\|\boldsymbol{X}_1 - \boldsymbol{X}_2\|] \neq 0$. By the continuous mapping theorem, $DDC_n(\boldsymbol{X},Y)$ converges to $DDC(\boldsymbol{X},Y)$ almost surely as $n \to \infty$.\\
\\
\noindent{(iv)} By the properties of multivariate normal distribution, \( X \mid Y = y \sim N(\rho y, 1 - \rho^2) \). Hence \( DDC(X, Y) \) is calculated as:
\[
DDC(X, Y) = 1 - \frac{E_{Y} \left[ E \left( \|X_1 - X_2\| \mid Y_1 = Y_2 = Y \right) \right]}{E \left( \|X_1 - X_2\| \right)} = 1 - \frac{\frac{2}{\sqrt{\pi}} \sqrt{1 - \rho^2}}{\frac{2}{\sqrt{\pi}}} = 1 - \sqrt{1 - \rho^2}.
\]
As the symmetry of \( X \) and \( Y \), \( DDC(Y, X) = DDC(X, Y) = 1 - \sqrt{1 - \rho^2}. \)
\end{proof}

\vspace{0.8em}  

\begin{proof}[\normalfont\bfseries Proof of Theorem~\ref{thm:asymptotic_normality}: The asymptotic normality of \( DDC_n(\boldsymbol{X},Y) \) under the independent hypothesis.]

	Before presenting the proof, we first introduce some definitions: \(\xi_{ik} = \|\boldsymbol{X}_i - \boldsymbol{X}_k\|\), \(\xi_i = E[\|\boldsymbol{X} - \boldsymbol{X}_i\|\mid \boldsymbol{X}_i]\), \(\xi = E[\|\boldsymbol{X}_1 - \boldsymbol{X}_2\|]\), and
	\[
	U_{n,i} = \frac{1}{\sqrt{n}(n-1)} \sum_{k=1}^{i-1} (\xi_{ik} - \xi_i - \xi_k + \xi)a_{ik},
	\]
	where
	\[
	a_{ik} = 
	\begin{cases}
		2 - n, & i = k + 1; \\
		2, & k \leq i - 2.
	\end{cases}
	\]
	Besides, for the samples \(\{(\boldsymbol{X}_i, Y_i)\}_{i=1}^n\), we assume that \(Y_1 \leq ... \leq Y_n\) throughout this proof. Therefore, the differential distance correlation takes the following form:
	\begin{equation}
		\label{eq:ddc_alternative_form}
		DDC_n(\boldsymbol{X},Y) = \frac{\dfrac{1}{n(n-1)}\left(2\sum\limits_{i>j}^n \|\boldsymbol{X}_i - \boldsymbol{X}_j\| - n\sum\limits_{i=1}^{n-1}\|\boldsymbol{X}_i - \boldsymbol{X}_{i+1}\|\right)}{\displaystyle\binom{n}{2}^{-1}\sum\limits_{i>j}^n \|\boldsymbol{X}_i - \boldsymbol{X}_j\|}.
	\end{equation}
	Denote the numerator in \eqref{eq:ddc_alternative_form} as \(R_n\), it is easy to verify that
	\begin{equation}
		\label{eq:Rn_martingale}
		\sqrt{n} R_n = \sum_{i=2}^n U_{n,i} + o_p(1).
	\end{equation}
	
	Let \(\mathcal{F}_i = \sigma \{\boldsymbol{X}_1, ..., \boldsymbol{X}_i\}\) be the \(\sigma\)-field generated by \(\{\boldsymbol{X}_k\}_{k \leq i}\). Obviously, \(E[U_{n,i} \mid \mathcal{F}_{i-1}] = 0\) for all \(i \geq 2\), which implies that \(\{U_{n,i}, \mathcal{F}_i\}\) is a mean-zero martingale sequence. Then, the martingale central limit theorem \citep{hall2014} will hold if
	\begin{equation}
		\label{eq:martingale_condition1}
		\sum_{i=2}^n E\left[U_{n,i}^2 \mid \mathcal{F}_{i-1}\right] \xrightarrow{P} (dVar(\boldsymbol{X}))^2,
	\end{equation}
	and for all \(\epsilon > 0\)
	\begin{equation}
		\label{eq:martingale_condition2}
		\sum_{i=2}^n E\left[U_{n,i}^2 I(|U_{n,i}| \geq \epsilon) \mid \mathcal{F}_{i-1}\right] \xrightarrow{P} 0.
	\end{equation}
	Denote
	\begin{align*}
		V_{jk} &= E[(\xi_{ij} - \xi_i - \xi_j + \xi)(\xi_{ik} - \xi_i - \xi_k + \xi) \mid \mathcal{F}_{i-1}], \\
		\sum_{i=2}^n E(U_{n,i}^2 \mid \mathcal{F}_{i-1}) &= \frac{1}{n(n-1)^2} \left\{ 4 \sum_{i>j}^{n-1} (n-2i) V_{ij} + \sum_{i=1}^{n-1} (n^2 - 4i) V_{ii} \right\} = 4 C_{n,1} + C_{n,2}.
	\end{align*}
	
	First, it is easy to demonstrate that \(E[C_{n,1}] = 0\) and
	\[
	Var(C_{n,1}) = \frac{1}{n^2(n-1)^4} \left[ \sum_{i>j}^{n-1} \sum_{k>l}^{n-1} (n-2i)(n-2k) E(V_{ij} V_{kl}) \right] = O(n^{-2}) E[V_{12}^2] \rightarrow 0.
	\]
	Hence, \(C_{n,1} \rightarrow 0\) in probability. Moreover, \(E[C_{n,2}] = (1 + O(n^{-1})) (dVar(\boldsymbol{X}))^2 \rightarrow (dVar(\boldsymbol{X}))^2\), and
	\[
	Var\left(C_{n,2}\right) \leq \frac{1}{n^2(n-1)^4}\sum_{i=1}^{n-1}E[(n^2-4i)^2V_{ii}^2] = O(n^{-1})E[V_{11}^2]\to 0,
	\]
	which implies \( C_{n,2} \to (dVar(\boldsymbol{X}))^2 \) in probability. Thus, \eqref{eq:martingale_condition1} holds. 
	
	Next, we show \eqref{eq:martingale_condition2} holds. By the Markov inequality,
	\[
	\sum_{i=2}^{n}E[U_{n,i}^2 I(|U_{n,i}| \geq \epsilon) \mid \mathcal{F}_{i-1}] \leq \frac{1}{\epsilon^2}\sum_{i=2}^{n}E[U_{n,i}^4 \mid \mathcal{F}_{i-1}].
	\]
	Therefore, it is sufficient to prove \(\sum_{i=2}^{n}E[U_{n,i}^4]\to 0\) by the law of large numbers. The derivation process is shown below:
	\begin{align*}
		\sum_{i=2}^{n}E[U_{n,i}^4] &= \frac{1}{n^2(n-1)^4}\sum_{i=2}^{n}E\left[\left(\sum_{j=1}^{i-1}(\xi_{ij}-\xi_i-\xi_j+\xi)a_{ij}\right)^4\right] 
		= \frac{1}{n^2(n-1)^4}\sum_{i=2}^{n}\left[E\left[\sum_{j=1}^{i-1}a_{ij}^4(\xi_{ij}-\xi_i-\xi_j+\xi)^4\right]\right. \\
		&\quad \left. + E\left[\sum_{j\neq k}^{i-1}a_{ij}^2a_{ik}^2(\xi_{ij}-\xi_i-\xi_j+\xi)^2(\xi_{ik}-\xi_i-\xi_k+\xi)^2\right]\right] 
		= O(n^{-1})\to 0.
	\end{align*}
	Hence, \eqref{eq:martingale_condition2} is proved. 
	
	Combining the result of martingale central limit theorem and \eqref{eq:Rn_martingale}, we have
	\[
	\sqrt{n}R_n \xrightarrow{D} N\left(0,(dVar(\boldsymbol{X}))^2\right).
	\]
	As \(\hat{\Delta}_n \to \Delta\) in probability, according to Slutsky's theorem,
	\[
	\sqrt{n}DDC_n(\boldsymbol{X},Y) = \sqrt{n}\frac{R_n}{\hat{\Delta}_n} \xrightarrow{D} N\left(0,\frac{(dVar(\boldsymbol{X}))^2}{\Delta^2}\right).
	\]
	The proof is complete.
\end{proof}

\begin{proof}[\normalfont\bfseries Proof of the two cases in Section~\ref{subsec:asymptotic_normality}.] 
	\mbox{}\\[0.1em]  
	 \noindent{(i)} \textit{\( X \) follows univariate normal distribution.} By the definition of \( dVar(X) \) and \( \Delta \), it is sufficient to demonstrate that \( \sigma^2 = \pi/3 - \sqrt{3} + 1 \) when \( X \sim N(0,1) \). By Theorem~7 in \citet{szekely2007}, if \( X \sim N(0,1) \), \( (dVar(X))^2 = 4(\pi/3 - \sqrt{3} + 1)/\pi \). Besides, \( \Delta = 2/\sqrt{\pi} \). Therefore, the proof is complete.
	
	\vspace{0.5em}
	
	\noindent{(ii)} \textit{\( X \) follows uniform distribution.} It suffices to prove \( \sigma^2 = 2/5 \) when \( X \sim U(0,1) \). Hence, some critical statistics of \( X \) are calculated as follows:
	\begin{align*}
		E\left(\|X_1 - X_2\|^2\right) = \int_{0}^{1} \int_{0}^{1} (x_1 - x_2)^2 \, dx_2 \, dx_1 = \frac{1}{6}, 
		E\left(\|X_1 - X_2\|\right) =  2 \int_{0}^{1} \int_{0}^{x_1} (x_1 - x_2) \, dx_2 \, dx_1  = \frac{1}{3}, \\
		E\left(\|X_1 - X_2\| \|X_1 - X_3\| \right) = E\left[ \left( E\left( \|X_1 - X_2\| \mid X_1 \right) \right)^2 \right] 
		= E\left[ \left( \left( X_1 - \frac{1}{2} \right)^2 + \frac{1}{4} \right)^2 \right] = \frac{7}{60}.
	\end{align*}
	By the expression of \( (dVar(X))^2 \) in \eqref{eq:distance_variance}, \( \sigma^2 = 2/5 \). The proof is complete.
	
\end{proof}

\begin{proof}[\normalfont\bfseries Proof of Proposition~\ref{prop:extension}: Properties of an alternative approach to DDC] 

	First, Lemma~3 is presented to establish an equivalent form of the independence between \( \boldsymbol{X} \in \mathbb{R}^p \) and \( Y \).
	
	\begin{lemma}
		\label{lem:monotone_independence}
		For any \( p \)-dimensional strictly monotone function vector \( \boldsymbol{m} = (m_1, \ldots, m_p) : \mathbb{R}^p \to \mathbb{R}^p \), \( \boldsymbol{X} \in \mathbb{R}^p \) and \( Y \) are independent is equivalent to \( \boldsymbol{m}(\boldsymbol{X}) \) and \( Y \) are independent.
	\end{lemma}
	
	\begin{proof}[Proof of Lemma~\ref{lem:monotone_independence}]
		Let the random vector \( \boldsymbol{X} = (X_1, \ldots, X_p) \), and \( \sigma (Z) \) be the \(\sigma\)-field generated by \( Z \). As \(\{m_i\}_{i=1}^p\) are strictly monotone, \( \sigma (X_i) = \sigma (m_i(X_i)) \) for each \( X_i \). Therefore,
		\[
		\sigma (\boldsymbol{X}) = \sigma \left( \bigcup_{i=1}^p \sigma (X_i) \right) = \sigma \left( \bigcup_{i=1}^p \sigma (m_i(X_i)) \right) = \sigma (\boldsymbol{m}(\boldsymbol{X})).
		\]
		Hence, the independence between \( \boldsymbol{X} \) and \( Y \) is equivalent to the independence between \( \boldsymbol{m}(\boldsymbol{X}) \) and \( Y \). The proof is complete.
	\end{proof}
	
	By Lemma~\ref{lem:monotone_independence}, it's natural that for strictly monotone function vector \( \boldsymbol{h} \) satisfying \( E(\|\boldsymbol{h}(\boldsymbol{X})\|) < +\infty \), \( DDC(\boldsymbol{h}(\boldsymbol{X}), Y) = 0 \) if and only if \( \boldsymbol{X} \) and \( Y \) are independent.
	
	Next, let's discuss the functional dependence property. By the proof of Theorem~\ref{thm:ddc_properties}, \( DDC(\boldsymbol{h}(\boldsymbol{X}), Y) = 1 \) indicates that \( \boldsymbol{h}(\boldsymbol{X}) = E(\boldsymbol{h}(\boldsymbol{X}) \mid Y) \) almost surely. Denote \( \boldsymbol{h}^{-1} = (h_1^{-1}, \ldots, h_p^{-1}) : \mathbb{R}^p \to \mathbb{R}^p \) as the inverse function vector of \( \boldsymbol{h} \), then \( \boldsymbol{X} = \boldsymbol{h}^{-1}[E(\boldsymbol{h}(\boldsymbol{X}) \mid Y)] \) almost surely. Besides, if \( \boldsymbol{X} \) can be determined by \( Y \) almost surely, then \( \boldsymbol{h}(\boldsymbol{X}) \) can be determined by \( Y \) almost surely, leading to \( E_{Y} [ E [ \|\boldsymbol{h}(\boldsymbol{X}_1) - \boldsymbol{h}(\boldsymbol{X}_2)\| \mid Y_1 = Y_2 = Y ] ] = 0 \) and thus \( DDC(\boldsymbol{h}(\boldsymbol{X}), Y) = 1 \). Moreover, following the proof of Theorems~\ref{thm:ddc_properties} and~\ref{thm:asymptotic_normality}, (iii) and (iv) of Proposition~\ref{prop:extension} are easily obtained. The proof is complete.
\end{proof}

\end{document}